\documentclass[reprint,superscriptaddress,nofootinbib,twocolumn]{revtex4-1}[12pt]
\usepackage{graphicx}
\usepackage{graphics}
\usepackage{amsmath}
\usepackage{amssymb}
\usepackage{dcolumn}

\newcommand{\aaps}{{Astron.~Astrophys.~Supp.}}

\newcommand{\nn}{\nonumber}
\newcommand{\sun}{\ensuremath{\odot}}
\newcommand{\kpc}{\ \text{kpc}}
\newcommand{\GeV}{\ \text{GeV}}
\newcommand{\TeV}{\ \text{TeV}}
\newcommand{\epp}{e^+e^-}

\begin{document}

\title{The Electron Injection Spectrum Determined by Anomalous 
Excesses in Cosmic Ray, Gamma Ray, and Microwave Data}
\author{Tongyan Lin}
\email{tongyan@physics.harvard.edu}
\author{Douglas Finkbeiner}
\affiliation{Harvard-Smithsonian Center for Astrophysics, 
60 Garden St., Cambridge, MA 02138, USA}
\affiliation{Physics Department, Harvard University, 
17 Oxford St., Cambridge, MA 02138, USA}
\author{Gregory Dobler}
\affiliation{Kavli Institute for Theoretical Physics,
Kohn Hall, Santa Barbara, CA 93106, USA}
\date{\today}
\begin{abstract}
Recent cosmic ray, gamma ray, and microwave signals observed by Fermi,
PAMELA, and WMAP indicate an unexpected primary source of $e^+e^-$ at
10-1000 GeV.  We fit these data to ``standard backgrounds'' plus a new
source, assumed to be a separable function of position and energy.
For the spatial part, we consider three cases: annihilating dark
matter, decaying dark matter, and pulsars.  In each case, we use
GALPROP to inject energy in log-spaced energy bins and compute the
expected cosmic-ray and photon signals for each bin. We then fit a
linear combination of energy bins, plus backgrounds, to the data. We
use a non-parametric fit, with no prior constraints on the spectrum
except smoothness and non-negativity.  In addition, we consider
arbitrary modifications to the energy spectrum of the ``ordinary''
primary source function, fixing its spatial part, finding this alone
to be inadequate to explain the PAMELA or WMAP signals.  We explore
variations in the fits due to choice of magnetic field, primary
electron injection index, spatial profiles, propagation parameters,
and fit regularization method. Dark matter annihilation fits well,
where our fit finds a mass of $\sim$1 TeV and a boost factor times
energy fraction of $\sim$70. While it is possible for dark matter
decay and pulsars to fit the data, unconventionally high magnetic
fields and radiation densities are required near the Galactic Center
to counter the relative shallowness of the assumed spatial
profiles. We also fit to linear combinations of these three scenarios,
though the fit is much less constrained.

\end{abstract}
\maketitle


\section{Introduction}
Several apparent anomalies in recent astrophysical data hint at a new
source of high energy electrons, positrons, and possibly gamma rays,
at the 10 GeV to 1 TeV scale.  The cosmic ray signals observed by
Fermi
\cite{Latronico:2009uw,Latronico:FermiSymposium,PesceRollins:2009af,Fermi1999}
and PAMELA \cite{Adriani:2008zr} are direct evidence for these
energetic electrons and positrons ($\epp$), which would lose their
energy primarily through synchrotron radiation and inverse Compton
scattering (IC).  If the number density of these $\epp$ rises towards
the Galactic Center (GC), then this synchrotron and IC could explain
the WMAP microwave ``haze" \cite{Dobler:2007wv} and the Fermi diffuse
gamma ray ``haze" \cite{Dobler:2009xz}, respectively.

It is difficult to explain these signals within the conventional
diffusive propagation model and with standard assumptions about the
interstellar medium (ISM). In this framework, the positron signal
arises from secondary production from spallation of proton cosmic rays
on the ISM. Assuming that 1. positrons and electrons have the same
energy losses, 2. primary electrons and protons have the same
production spectrum, and 3. the proton escape time decreases with
energy, then the predicted positron fraction generically falls with
energy, in contrast to the rising fraction observed by PAMELA.  Katz
et al. \cite{Katz:2009yd} point out these assumptions can be wrong,
and explore alternative scenarios. Indeed, secondary production at
shock fronts could explain the $e^+$ excess \cite{Blasi:2009hv,
Blasi:2009bd}, but this would also imply an excess of anti-protons,
which is not observed. We will not consider these alternatives further.

\begin{figure*}[t]
\begin{center}
\begin{align}
\text{(a)} & \includegraphics[width=.43\textwidth]{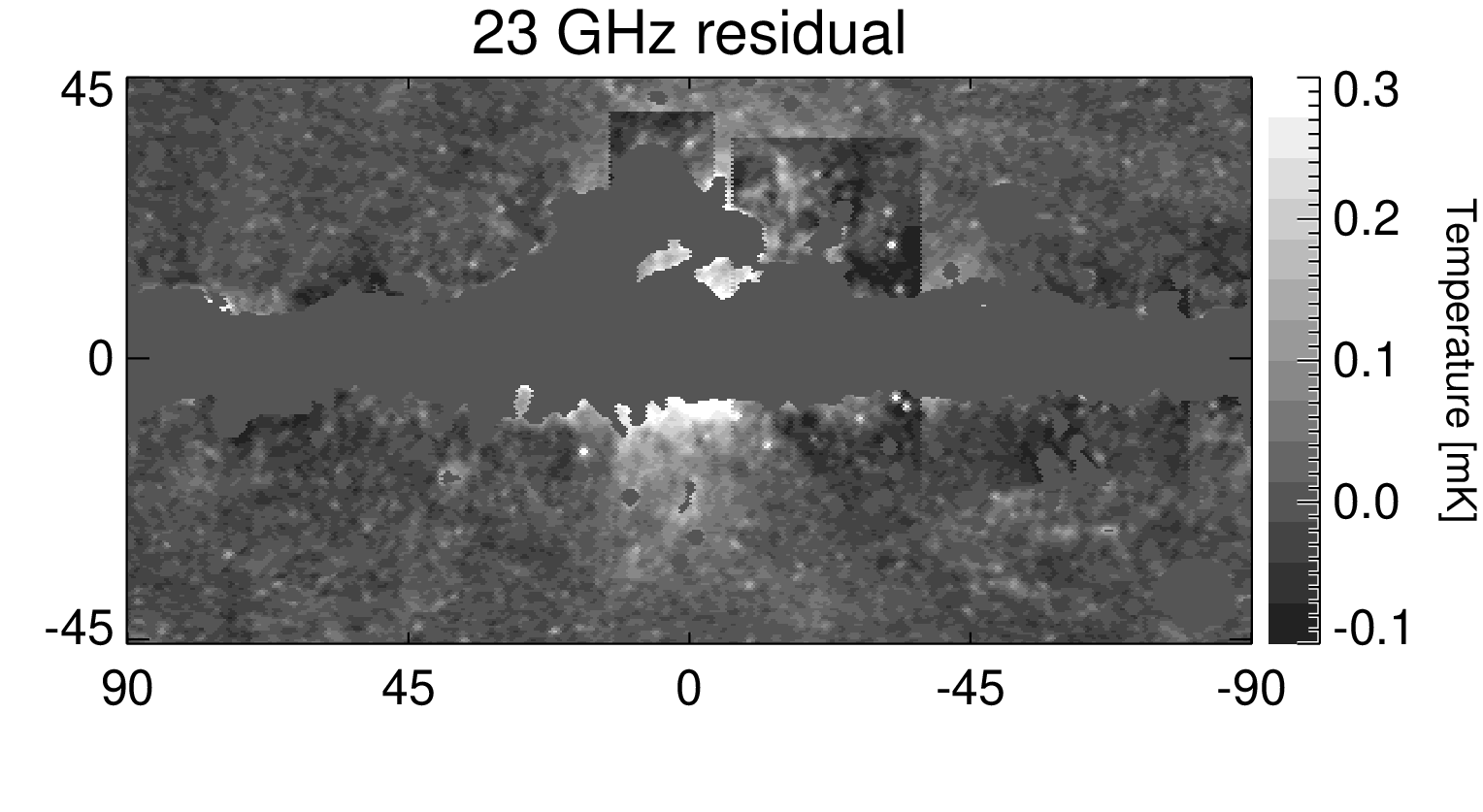}
\text{(b)} & \includegraphics[width=.43\textwidth]{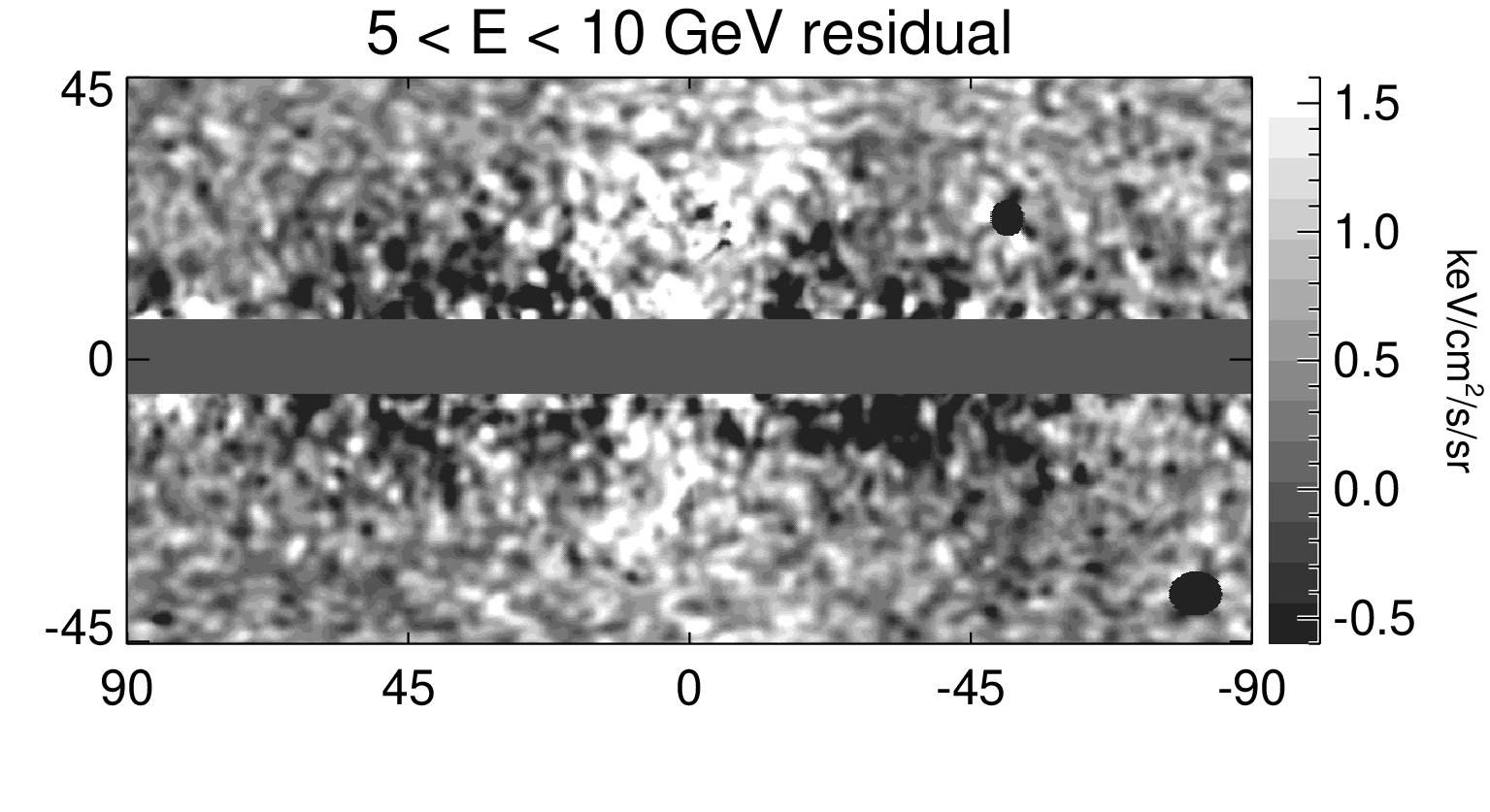} \nn \\
\text{(c)} & \includegraphics[width=.4\textwidth]{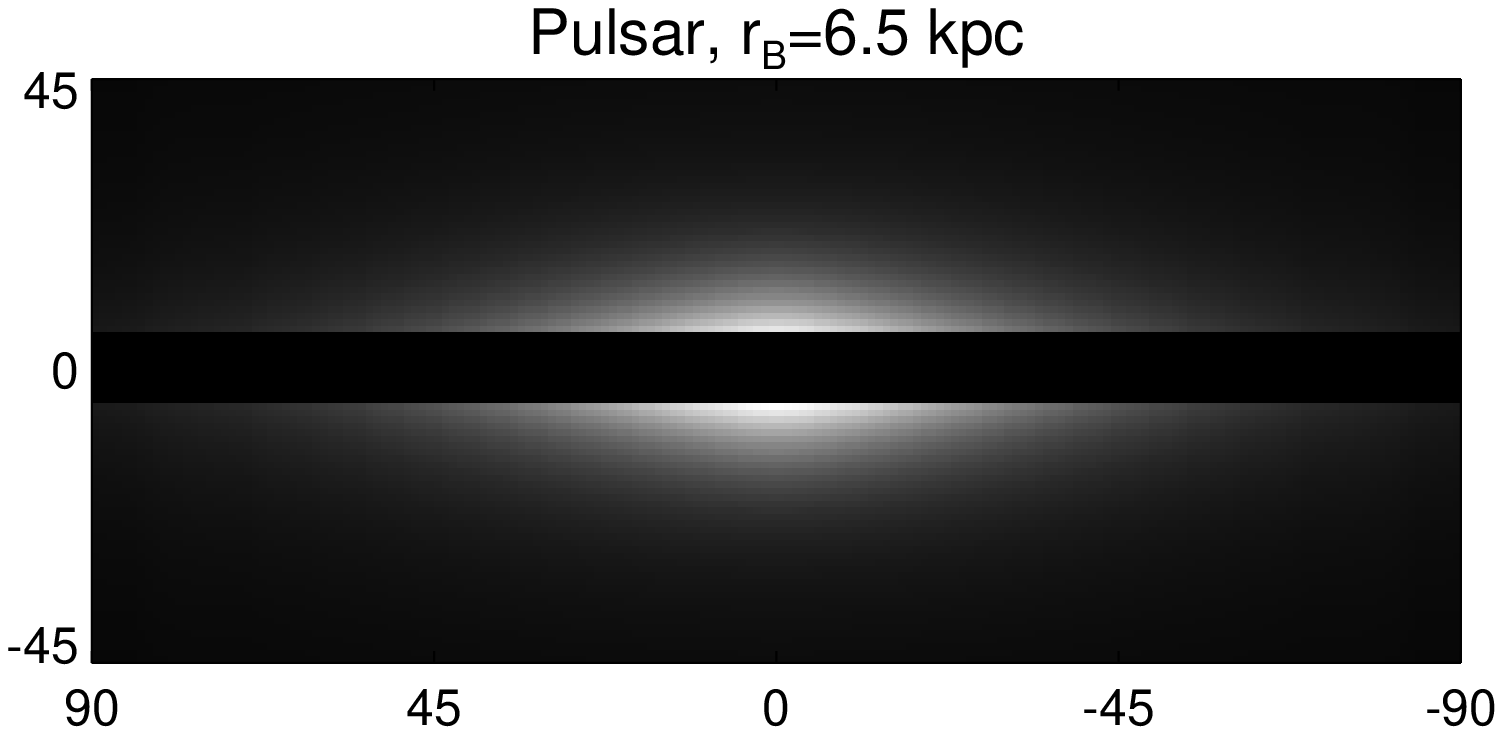} \hspace{.6cm} 
\text{(d)} & \includegraphics[width=.4\textwidth]{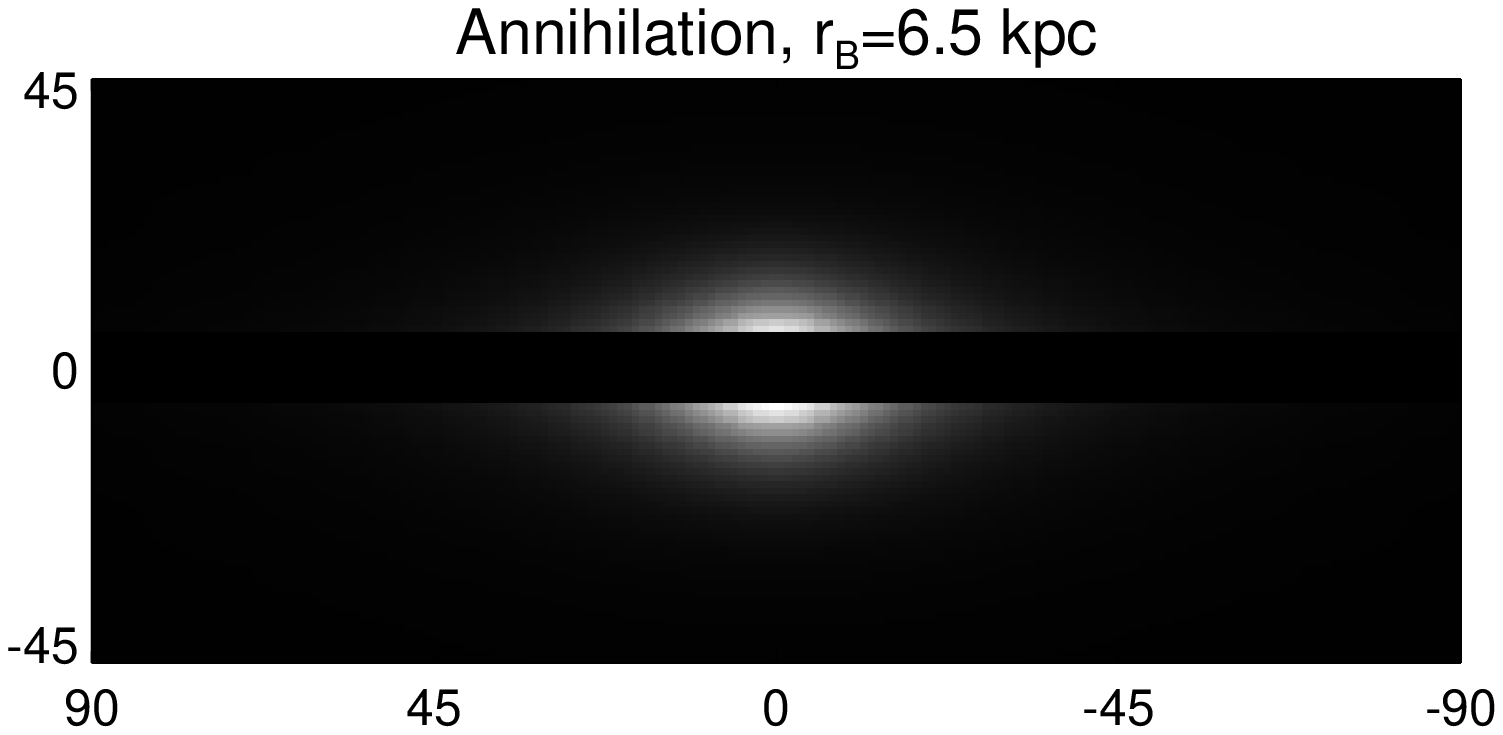} \hspace{.6cm} \nn \\
\text{(e)} & \includegraphics[width=.4\textwidth]{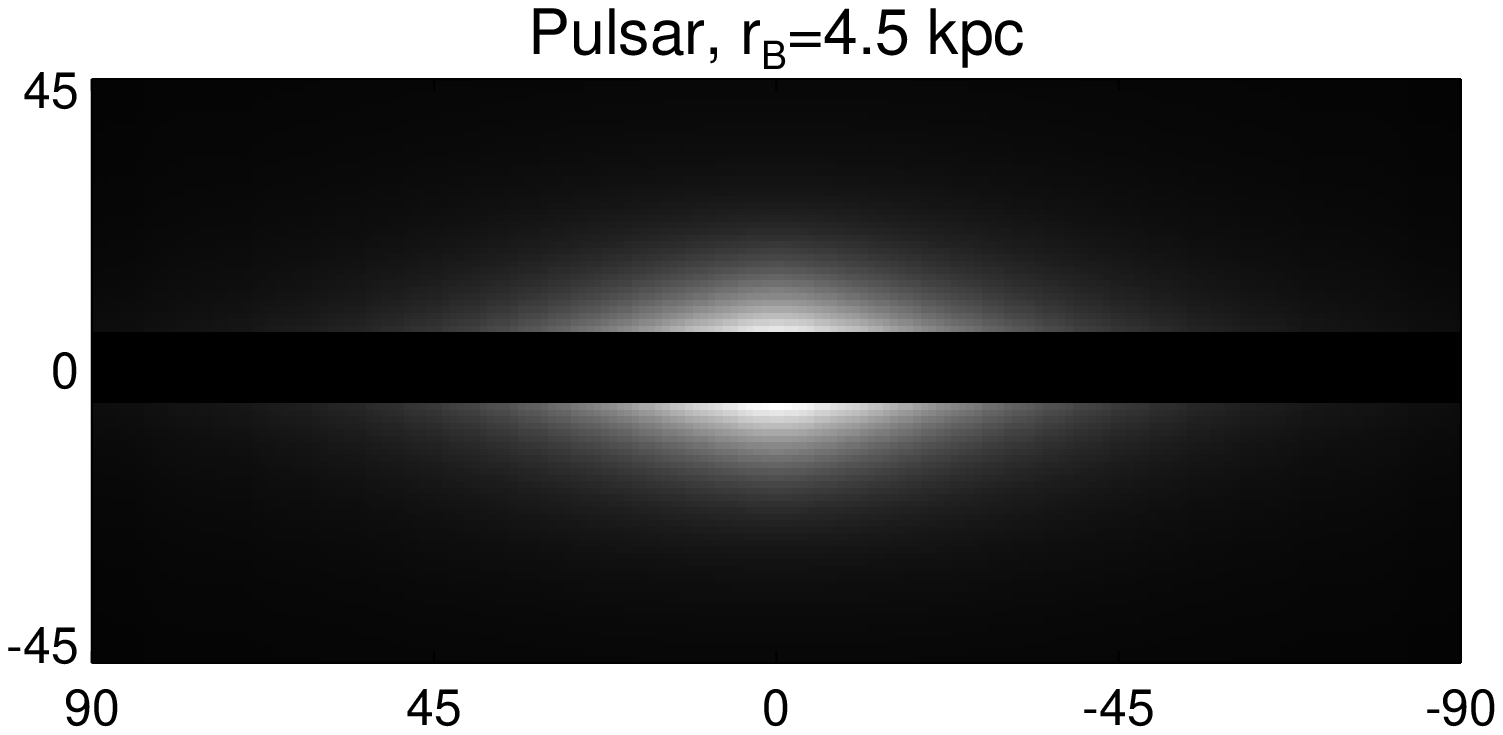} \hspace{.6cm}
\text{(f)} & \includegraphics[width=.4\textwidth]{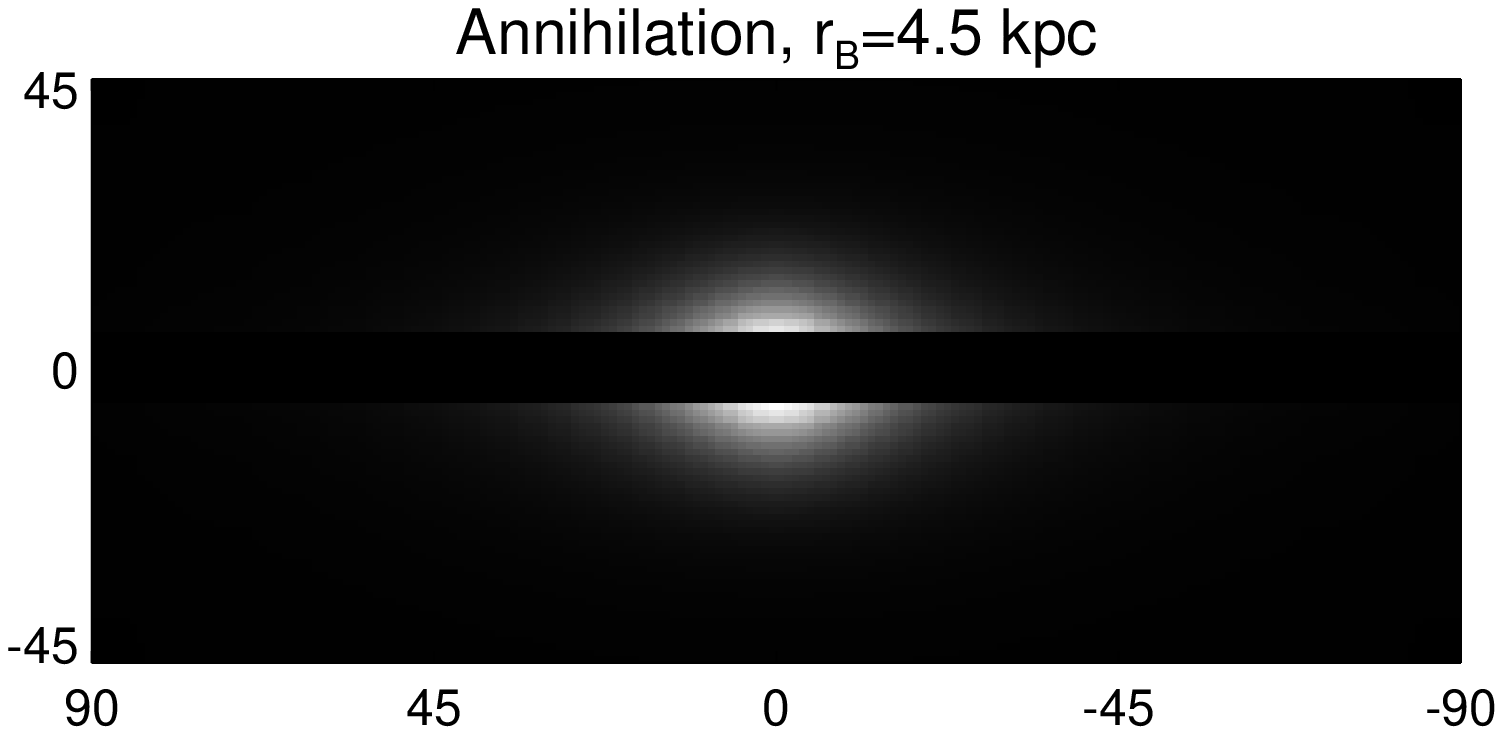} \hspace{.6cm}  \nn
\end{align}
\end{center}
\caption{Maps of the (a) WMAP haze at 23 GHz and (b) Fermi gamma-ray
haze at 5-10 GeV for the region $-90<\ell<90$ and $-45<b<45$. The
gamma-ray haze is obtained by subtracting the Fermi diffuse model from
the data. All maps are centered on the GC. The data are compared to
normalized maps of (c) pulsar synchrotron at 23 GHz and (d)
synchrotron at 23 GHz from dark matter annihilation with an Einasto
profile. The magnetic field has the form of Eq.~\ref{eq:Bfield} with
$r_B = $ 6.5 kpc. The morphology of the haze signals more closely
resembles the signals from dark matter than from pulsars.  We show the
corresponding results for $r_B = $ 4.5 kpc in (e) and (f). Choosing a
steeper magnetic field can change the morphology, but this is not
preferred by the Haslam data (see Fig.~\ref{fig:haslam}). The region
$-5 \le b \le 5$ is masked out because we only fit the region $b \le
-5$.}
\label{fig:haze}
\end{figure*}

We examine here whether a new primary source of $\epp$ is a viable
explanation of the signals.  First, the rise in the positron fraction
measured by PAMELA suggests the presence of a new hard source of
positrons \cite{Serpico:2008te}. Second, the WMAP ``haze'' is
consistent with a hard synchrotron signal in the inner galaxy, in
addition to a soft-spectrum synchrotron component traced by
Haslam. Though this decomposition is not unique, it is a good fit to
the WMAP data. Third, the Fermi gamma ray ``haze'' similarly extends
to $|b| > 30^\circ$ above and below the plane in the inner galaxy.
Neither haze correlates with the morphology of any known astrophysical
objects or the ISM. (See Fig.~\ref{fig:haze}.)

Many attempts to explain the data operate by including a new component
of high energy particles and gamma rays originating from one of the
following sources:
\begin{enumerate}
	\item Annihilation of TeV-scale dark matter, 
	\item Decay of TeV-scale dark matter, or
	\item An astrophysical source such as pulsars.
\end{enumerate}
These sources can produce energetic electrons, positrons, and gamma
rays.  In addition, the dark matter distribution in the Galaxy is
expected to be roughly spherical, providing at least qualitative
agreement with the morphology of the gamma-ray and microwave haze.
Nevertheless, each explanation above has drawbacks.

While annihilating dark matter may seem natural given a weak-scale
WIMP which has a thermal freeze-out annihilation cross section, this
vanilla scenario cannot explain the observed signals.  Boost factors
in the annihilation rate, arising from substructure or particle
physics enhancement, of order 100-1000 are typically needed, depending
on the annihilation channels and dark matter mass.  Significant
model-building effort is also required to explain the lack of excess
in the observed $\bar p/p$ flux \cite{Adriani:2008zq}.  For examples,
see
\cite{Cirelli:2008id,Cirelli:2008pk,ArkaniHamed:2008qn,Cholis:2008wq,
Mardon:2009rc}.

In the decaying dark matter scenario, dark matter has the freeze-out
annihilation cross section but also decays with lifetime $\tau_\chi
\sim 10^{26}$ s. These models also must explain why there is no excess
in $\bar p/p$, though no boost factors are required. Examples include
\cite{Nardi:2008ix, Mardon:2009gw, Arvanitaki:2008hq, Ibarra:2009dr,
Ruderman:2009ta, Chen:2008yi, Yin:2008bs}.
 
The pulsar explanation is the least exotic, but there are significant
astrophysical uncertainties in pulsar distributions and $e^+e^-$
emission spectra.  The Fermi cosmic ray signals can be explained by
the presence of one or more nearby pulsars with hard $\epp$ emission
spectra \cite{Hooper:2008kg, Profumo:2008ms, Yuksel:2008rf,
Malyshev:2009tw, Gendelev:2010fd}. However, pulsars are generally
expected to be concentrated in the disk and it can be difficult to
explain the shape of the WMAP and Fermi ``haze" signals, which are
much more spherical.  See also \cite{Kaplinghat:2009ix,
Harding:2009ye}.

In this paper we quantify how well each of these three scenarios can
explain the data described above without resorting to model-dependent
details of the particle physics or pulsars.  Rather we use the data to
determine the best-fit injection spectrum of electrons and positrons
produced by each new source. We also show that a simple modification
to the background electron injection can explain the Fermi $e^+ + e^-$
spectrum and the Fermi gamma ray spectrum but not the rest of the
data.


\begin{table}[tb]
\begin{center}
\begin{tabular}{|c|c|c|c|}
\hline 
 & $K_0 \ [\kpc^2/\text{Myr}]$  & $\delta$ & L [kpc] \\
\hline
Default\ & 0.097 & \ 0.43  \ & 4 \\
M1	&  0.0765 & \  0.46 \  & 15 \\
MED & 0.0112 & \ 0.70  \ & 4 \\
M2 & 0.00595 &  \ 0.55  \ & 1  \\
\hline
\end{tabular}
\end{center}
\caption{Typical propagation parameters consistent with low-energy
cosmic ray data \cite{Delahaye:2007fr}.  We use the ``Default"
parameters and show the effect of using M1 and MED in
Fig~\ref{fig:errors}.}
\label{tab:prop}
\end{table}


The standard procedure to analyse whether a model can explain the
astrophysical signals is:
\begin{center}
	pulsar or particle physics model \\
	$\Downarrow	$ \\
	spectrum of particles produced by the source\\
	$\Downarrow	$ \\
	propagation (e.g., GALPROP) \\
	$\Downarrow	$ \\
	comparison with data
\end{center}
Often, one fits a specific dark matter or pulsar model to a subset of
the ``anomalous'' signals described above. For dark matter, the
particle physics model is usually processed through Pythia
\cite{Sjostrand:2006za} to generate a spectrum of $\epp$.  The
injection spectrum is the spectrum of $\epp$ produced per unit source
times the rate of production of $\epp$ per source and the spatial
distribution of the source. These injected $\epp$ are propagated
through the Galaxy to obtain a steady-state solution.  The signals are
then compared with data.

While some analyses have studied the cases above in a less
model-dependent way, the injection spectrum is assumed to have one of
a few common forms \cite{Cirelli:2008pk,Barger:2009yt,Zhang:2009ut}.


In this paper we effectively reverse the arrows in the procedure
above. We fit the data from Fermi, PAMELA, and WMAP to expected
backgrounds plus a new source which produces positrons and
electrons. We assume the injection of the new source is separable in
position and space. Rather than specifying the spectrum of $e^+e^-$
injected by the new source, we fit for this injection spectrum as a
function of energy. The shape of the spectrum is the same everywhere,
and the spatial distribution is varied over several conventional
models. Therefore, for our purposes, the three scenarios listed above
differ only in their spatial distributions.

In the pulsar case, the assumption that the pulsar injection is a
separable function of position and energy is a crude approximation that
allows us to fit the data without specifying the details of pulsar
physics, since the position dependence of pulsar populations and their
$\epp$ injection spectra is very uncertain.

In our fits of the injection spectra, we simultaneously account for
possible variations in the conventionally assumed spectrum and spatial
distribution of the background injected electrons, as well as
propagation parameters, magnetic field, and starlight densities. This
takes into account the uncertainties in current models of the Galactic
backgrounds.

We describe the signals and their expected backgrounds in more detail
in the next section.  We then present the overall framework of the
analysis. Predictions are computed using GALPROP, and we allow for
variations in the background model.  We then present the best-fit
injection spectrum for each of the three scenarios above, as well as
the best-fit of the data to an arbitrary modification of the
background electron injection spectrum. Finally we present injection
spectra for linear combinations of these scenarios.


\section{Signals \label{sec:signals}}

In this section we review the method of computing the signals
and standard assumptions made in modeling the astrophysical
backgrounds. However, in our fits we allow for variations in many of
these assumptions. This is discussed in more detail in
Sec.~\ref{sec:uncertainties}.

In the conventional diffusive propagation model, the $e^-$ cosmic
ray density, $dn(\vec x, E)/dE$, is the steady-state solution to the 
diffusion and energy loss equation:
\begin{align}
\label{eq:diffusion}
  \frac{\partial}{\partial t} & \left( \frac{dn(\vec x, E)}{dE}\right)  =  0 \\
 = & \vec\nabla \cdot \left[ K(E) \vec\nabla \frac{dn}{dE}\right] + 
\frac{\partial}{\partial E} \left[ b(E,\vec x) \frac{dn}{dE} \right] + Q(E,\vec x) \nn
\end{align}
where the first term represents diffusion, the second term energy
loss, and the third term the source term. $K(E)$ is the diffusion
coefficient and $b(E, \vec x)$ is the energy loss rate. This equation
holds separately for positrons. For both electrons and positrons,
diffusive re-acceleration and galactic convection are negligible above
a few GeV. Those effects are often relevant for other cosmic rays,
which are governed by similar equations.  We use GALPROP v50p.1 to
solve for steady state cosmic ray densities. For a review, see
\cite{Strong:2007nh}.

For electrons, the source term includes primary electrons produced by
supernovae and secondary electrons produced by collisions of proton
cosmic rays on the ISM. We denote these sources by $Q_0(E, \vec
x)$. The source term can also include any new source of electrons,
$Q_1(E, \vec x)$. For positrons, the source term includes only
secondary positrons and any new source of positrons. The spectrum of
injected secondary $\epp$ is determined by the astrophysics of proton
cosmic rays and their interactions.

The injected primary electron spectrum is usually assumed to have the
following energy dependence:
\begin{align}
	\label{eq:primary_electrons}
	\frac{dN}{dE} \propto
	\begin{cases}
	E^{1.6} & , E < 4 \GeV \\
	E^{\gamma_e} & , 4 \GeV<E < 2.2 \TeV  \\
	E^{3.3} & , E > 2.2 \TeV 		 
	\end{cases}
\end{align} 
where $\gamma_e$ can vary. $dN/dE$ is the spectrum of $e^-$ per unit
source and is continuous.  Eq.~\ref{eq:primary_electrons} has often
been adopted in the past because the resulting cosmic ray fluxes
approximately agreed with the available data. Though we use this form
as a default, we will also fit for an arbitrary modification to
$dN/dE$.

The number density for the supernovae that inject these
electrons is commonly parametrized as
\begin{equation}
	n_s(\vec x) \propto r^\alpha 
	\exp \left( -\beta \frac{ r}{r_\sun} - \frac{|z|}{.2 \kpc} \right) \Theta(r_{max} - r)
	\label{eq:source_dist}
\end{equation} 
where $r$ is distance to the center of Galaxy, projected on the
galactic plane, and $z$ is distance perpendicular to the galactic
plane. The default GALPROP parameters are $\alpha = 2.35, \beta =
5.56283,$ and $r_{max} = 15 \kpc$ \cite{Strong:2004td,Lorimer:2003qc}.

The default normalization of the product $n_s \times dN/dE$ is fixed such
that the observed local flux from the primary electrons satisfies
\begin{align}
	\frac{c}{4\pi} \frac{dn}{dE} & (34.5 \GeV, z=0, r=r_\sun) = \nn \\
	& 3.15922 \times 10^{-7} (\text{cm}^2 \cdot \text{sr} \cdot \text{s} \GeV)^{-1}
	\label{eq:primary_normalization}
\end{align}
which is consistent with the flux observed by Fermi.

The diffusion of the injected $\epp$ is governed by the diffusion
coefficient, $K(E)$, and $L$, the escape distance out of the galactic
plane.  $K(E)$ represents the random walk of a charged particle in
tangled magnetic fields, and is approximated as constant in space. It
is generally assumed that $K(E) = K_0 (E/ \text{GeV})^\delta$, where
$K_0$ and $\delta$ are propagation parameters.  In
Table~\ref{tab:prop} we give some commonly used values of $K_0$,
$\delta$, and $L$.  \cite{Delahaye:2007fr,Cumberbatch:2010ii}.  Our
default model assumes $K_0 = 0.097 \kpc^2/\text{Myr}$, $\delta =
0.43$, and $L = 4 \kpc$, though we will vary these parameters later.
This choice matches cosmic ray data for protons, the B/C cosmic ray
ratio, and was used in \cite{Cholis:2008hb}.


\begin{figure*}[thb]
\begin{center}
(a)\ \includegraphics[width=.45\textwidth]{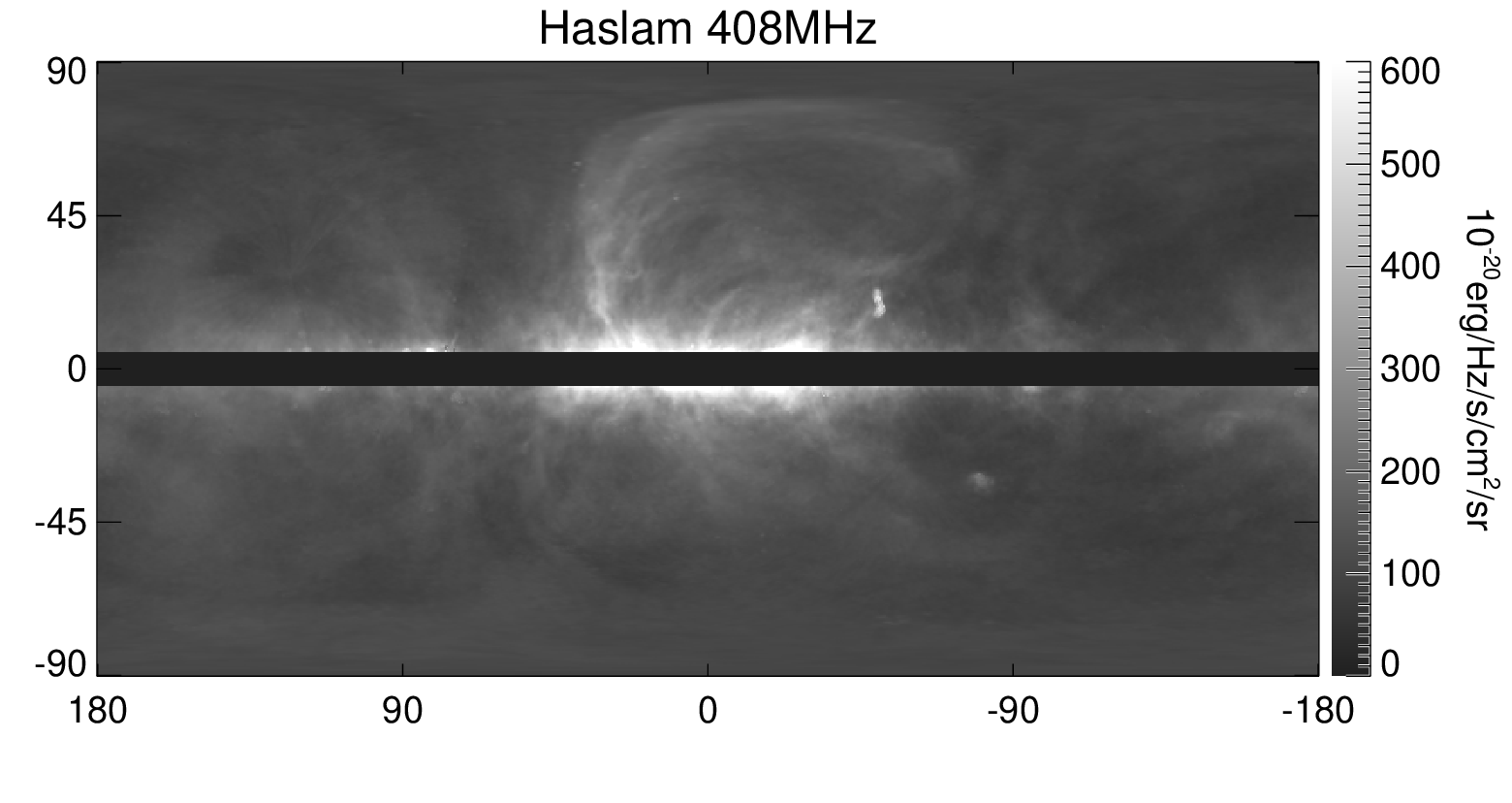}
(b)\ \includegraphics[width=.45\textwidth]{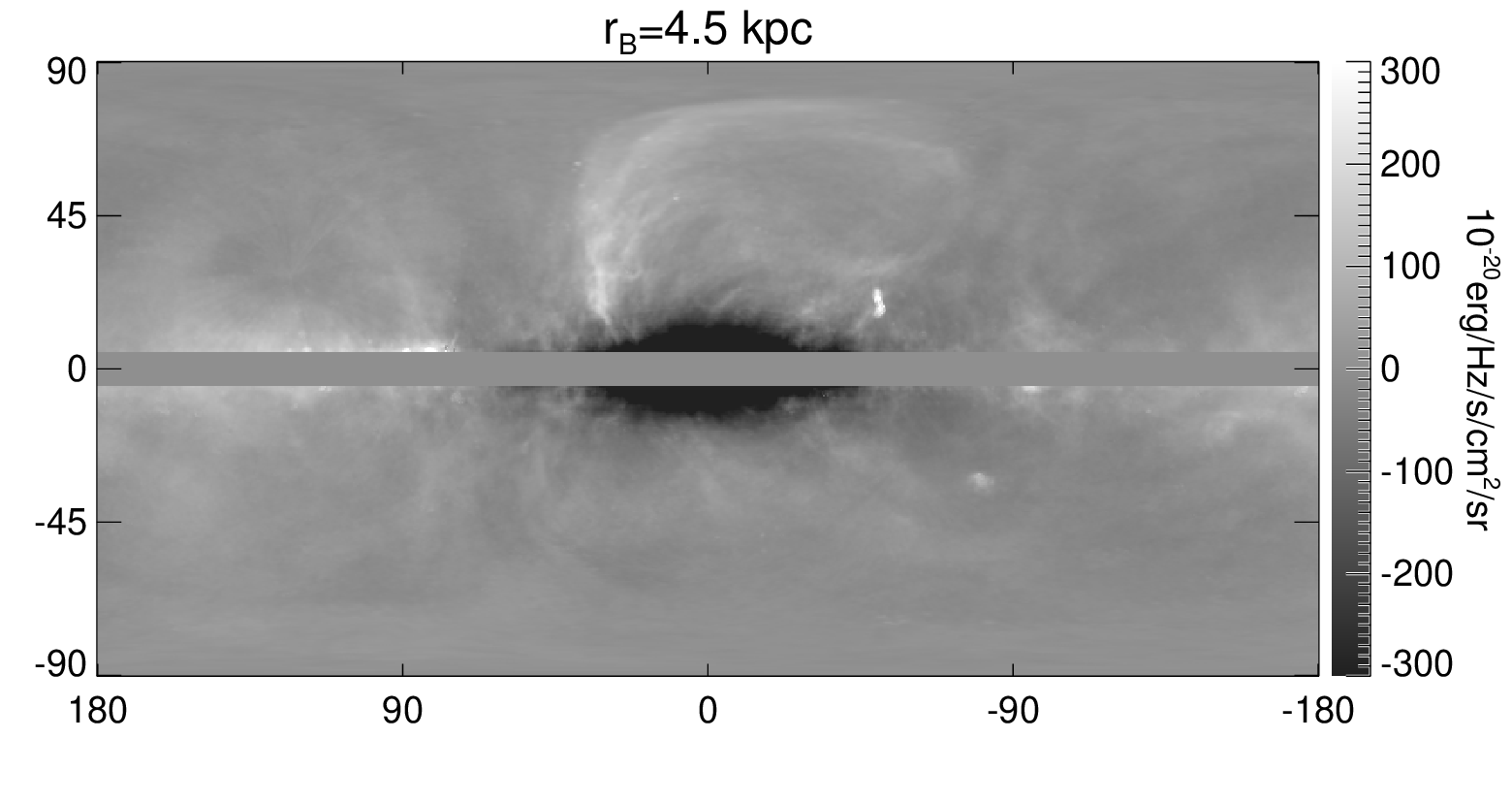} \\
(c)\ \includegraphics[width=.45\textwidth]{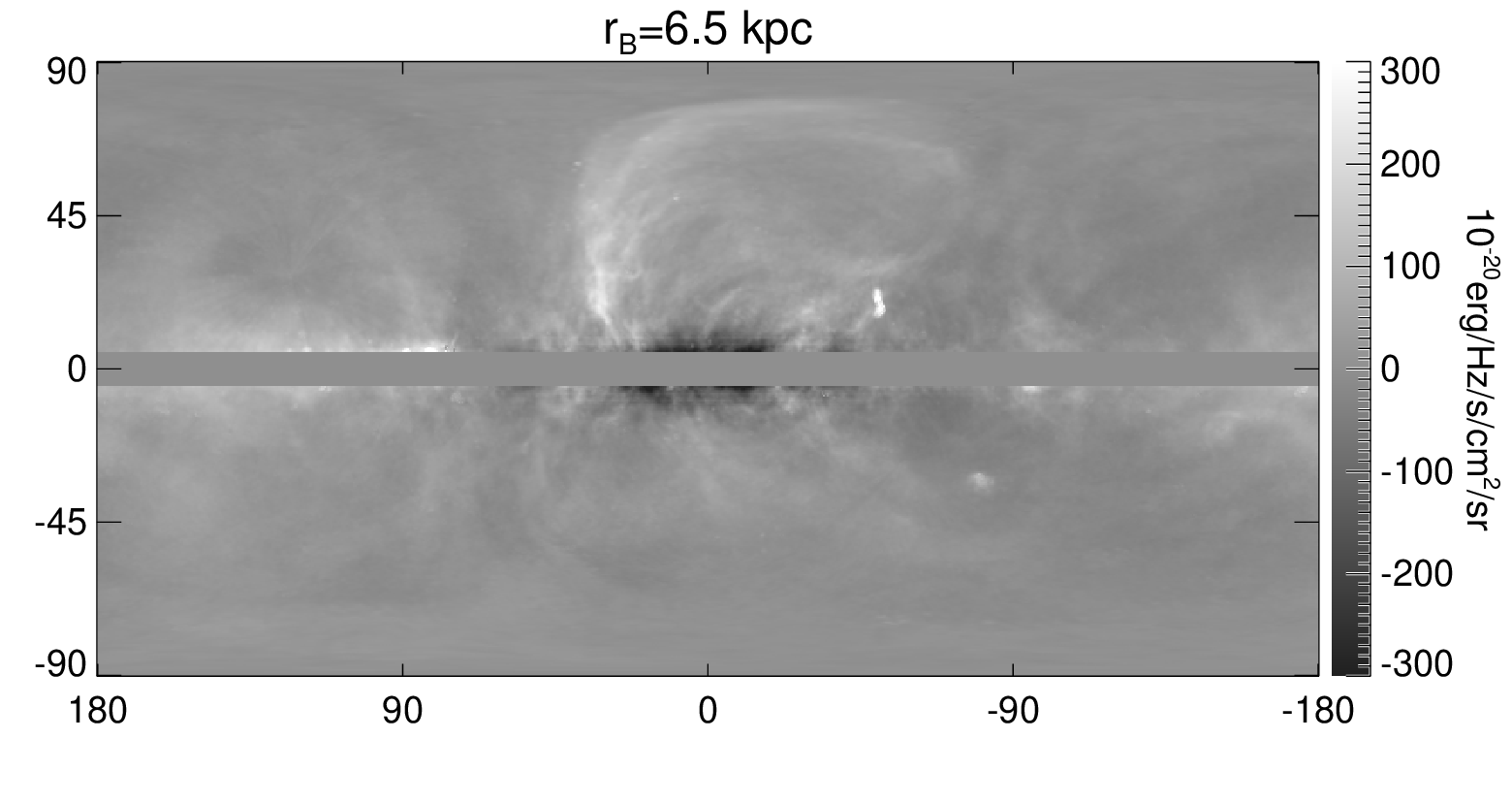}
(d)\ \includegraphics[width=.45\textwidth]{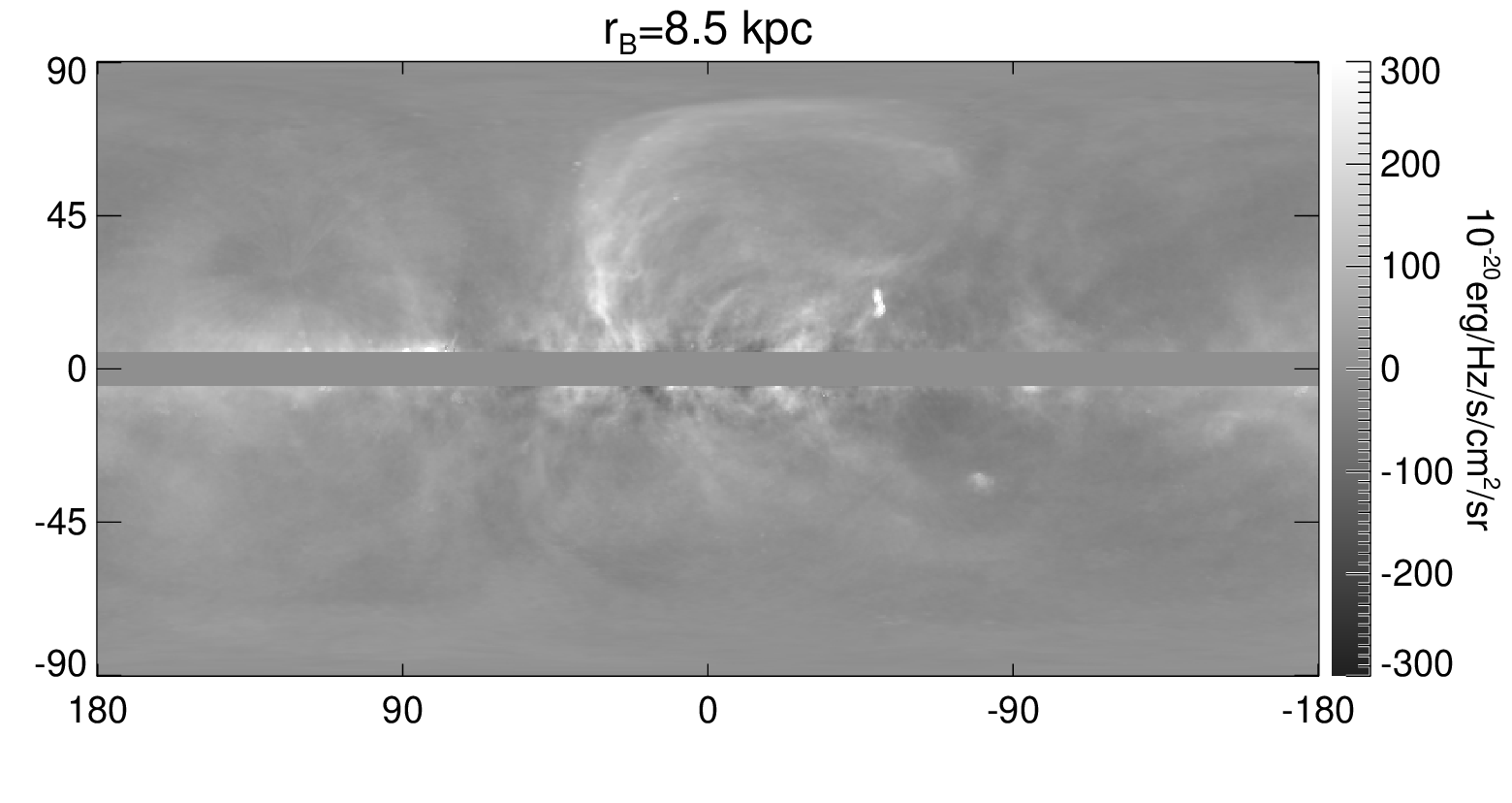}
\end{center}
\caption{(a) Haslam 408 MHz map. In the other panels we subtract the
default GALPROP model for (b) $r_B = 4.5 \kpc$, (c) $r_B = 6.5 \kpc$,
and (d) $r_B = 8.5 \kpc$ from the Haslam map. The GALPROP model is
normalized such that the total emission in the region $\ell\in
[-10,10], b \in [-90,-5]$ matches the Haslam 408 MHz intensity in the
same region. The constant offset is subtracted from the Haslam 408 MHz
data.  Note that local features like the North polar spur are not
modeled in GALPROP and hence are not fit.}
\label{fig:haslam}
\end{figure*}


As injected electrons and positrons propagate, they lose energy.  The
energy loss rate $b(E, \vec x)$ includes energy loss mechanisms.  The
path length for an electron or positron losing $1/e$ of its energy is
typically given by $\sim \sqrt{K E/b}$, which is $\sim 1 \kpc$ around
1 GeV and becomes shorter for higher energies, at least until the
Klein-Nishina limit \cite{Stawarz:2009ig}.  The dominant mechanisms
for energy loss are IC scattering and synchrotron, where $b(E) \propto
E^2$.  Bremsstrahlung (brem), for which $b(E) \propto E$, is
sub-dominant above $\sim 1 \GeV$ and is far more localized to the
disk. For a new high energy component of $e^+e^-$, we neglect
bremsstrahlung.

The IC rate depends strongly on the model for the interstellar
radiation field.  We use the default GALPROP model
\cite{Moskalenko:2005ng} as a baseline.  Meanwhile, the synchrotron
depends on the model for the magnetic field.  We assume a standard
parametrization of the field,
\begin{equation}
	|B| = B_0 \exp\left(-\frac{r-r_\sun}{r_B}\right)  
                  \exp\left(-\frac{|z|}{z_B}\right).
	\label{eq:Bfield}
\end{equation} 
$r$ is the distance to the center of the Galaxy, projected on the
galactic plane. Typical parameters are $B_0 = 5 \mu$G, $r_B \sim 5-10
\kpc$, and $z_B = 2 \kpc$. \footnote{The documentation for GALPROP
v50p incorrectly states that their parameter $B_0$ is the magnetic
field in the center of the galaxy.}  For our default propagation
parameters, the GALPROP synchrotron prediction at 408 MHz best matches
the Haslam 408 MHz map if $r_B \approx 8.5 \kpc$; see
Fig.~\ref{fig:haslam}.

This parameterization is consistent with observations of the large-scale
(ordered) magnetic fields at 1-10 kpc \cite{Han:2006ci}. The random
component of the magnetic field is assumed to be proportional to the
ordered fields, with a proportionality factor of approximately one
\cite{Jaffe:2009hh}. Thus Eq.~\ref{eq:Bfield} is sufficient for our
purposes, since our fits are not sensitive to the detailed structure
of the magnetic fields. We increase or decrease the average strength
of the magnetic fields in the Galactic Center region by decreasing or
increasing $r_B$.

The solution to Eq.~\ref{eq:diffusion} is the steady-state cosmic ray
density, which then determines the photon signals. The gamma ray flux
includes decay of $\pi^0$s produced in proton cosmic ray collisions
with the gas in the ISM, IC scattering of $e^\pm$ on interstellar
photons, and bremsstrahlung of $e^\pm$ colliding with the ISM. The
gamma-ray power in a given direction scales as:
\begin{align}
	P_{\pi^0} & \propto \int n_{gas}( s)\ n_p(s)\ ds \ , \\
	P_{\text{IC}} & \propto  \int n_*( s)\ n_{e^\pm}(s)\ ds \ , \\
	P_{\text{brem}} & \propto \int n_{gas}( s)\ n_{e^\pm}(s)\ ds.
\end{align} where $s$ is the coordinate along the line of sight.
The $\sim$23 GHz microwave flux off the Galactic plane is primarily
synchrotron radiation of electrons and positrons
\begin{equation}
	P_{\text{synch}} \propto \int |B(s)|^2 \ n_{e^\pm}(s)\ ds
\end{equation}  where $B$ is the magnetic field.

A new source such as dark matter or pulsars can inject high energy
electrons and positrons at 10 GeV to 1 TeV. These new sources are
included in $Q_1(E,\vec x)$. In this paper we solve for the $\epp$
injection spectrum which, after propagation, yields the observed
cosmic ray spectrum and gives rise to gamma rays and synchrotron
radiation. Our fit will essentially determine $Q_1(E, \vec x_0)$, where
$\vec x_0$ is the Earth's location. The spatial dependence of $Q$ is
fixed to be one of a few conventional models.

These sources can also directly inject photons. There are primary
photons from pulsars which are important at lower gamma-ray energies.
Given our energy range of interest, we do not consider these further.

In the case of dark matter annihilation or decay, generally there are
many channels through which dark matter produces Standard Model
particles, which can then decay on short time scales. The end products
are $e^{\pm}$, neutrinos, and photons.  However, we do not consider
these direct gamma rays any further. These can be produced from
$\pi^0$s, final state radiation\footnote{In some papers, final state
radiation is referred to as internal bremsstrahlung. We use
``bremsstrahlung'' exclusively to mean $\epp$ cosmic rays colliding
with the ISM.} from $\tau^{\pm}$s or $\mu^{\pm}$s, or a direct photon
channel. For TeV-scale dark matter, these gamma rays can have higher
energies than those observed by Fermi. Furthermore, in the fits
below it is not difficult to produce enough gamma ray signal above
10-100 GeV. In fact, direct gamma ray production can be rather
constrained by observations \cite{Cirelli:2009vg,Papucci:2009gd}.


\subsection{Data \label{sec:data}}

We fit to the following data:
\begin{itemize}
	\item PAMELA $J(e^+)/(J(e^-) + J(e^+))$ positron
          fraction, which displays a steep rise from 10-100 GeV
          \cite{Adriani:2008zr}
	\item Fermi $(e^+ +e^-)$ cosmic ray spectrum, which shows a
          slight hardening of the spectrum at a few hundred GeV
          \cite{Latronico:2009uw,Latronico:FermiSymposium,PesceRollins:2009af}
	\item Fermi gamma ray spectrum, which shows a hardening of the
          spectrum at around 10-100 GeV, averaged over the haze region
          $-15<\ell <15$ and $10< |b| < 30$.  Note the pion signal
          has been subtracted from the data \cite{Dobler:2009xz}.  Our
          background models match the pion component, shown in Fig.~11
          of \cite{Dobler:2009xz}. This is not affected by the
          inclusion of new sources of electrons.
	\item WMAP synchrotron at 23, 33, and 41 GHz averaged over
          $-10<\ell <10$ for $-90< b < -5$, in $2$ degree bins. We
          also fit to the same data averaged over $10< |\ell| <30$,
          which we call the ``high $\ell$" region of the WMAP data and
          is incorporated to include morphological information from
          the microwave haze. \cite{Dobler:2007wv}
\end{itemize}
These data describe the ``anomalous signals", which suggest the
presence of a new source of electrons and positrons, and possibly
gamma rays, at roughly 10-1000 GeV.

The Fermi LAT collaboration has provided a reference model for the
diffuse emission \cite{FermIDiffuseModel}, a detailed fit that
includes a reference GALPROP model for IC and models for a number of
residual local features giving rise to bremsstrahlung at lower
energies.  Since we are not studying the detailed structure of the
diffuse gamma rays and because IC and pions dominate at high energies,
it is sufficient for us to use GALPROP to model the diffuse gamma ray
emission in the haze region.

\begin{figure}[t]
\begin{center}
\includegraphics[width=.45\textwidth]{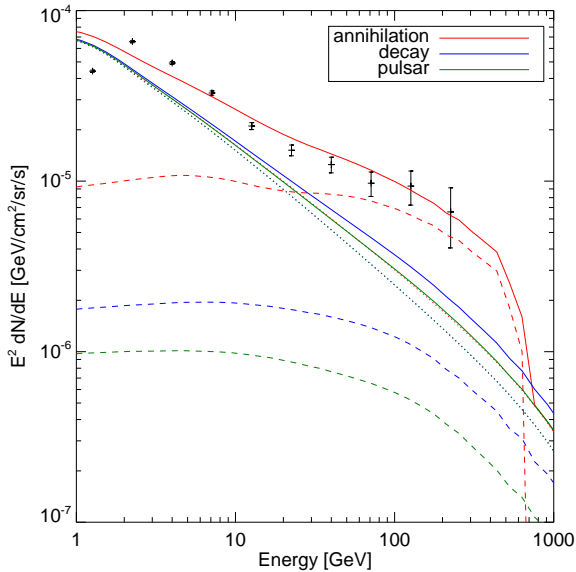}
\end{center}
\caption{ The fits in this paper are consistent with Fermi
  observations of the gamma ray spectrum in the region $ |\ell| \le
  3$, $|b| \le 3$. Point sources have been subtracted.  Solid colored
  lines show the predicted total signal for the best fits of the new
  sources considered in this paper, including backgrounds. The dashed
  lines show the contribution from only the new source. }
\label{fig:gammas}
\end{figure}

We also do not attempt to fit the Fermi gamma-ray spectrum near the
Galactic Center region nor the Fermi isotropic gamma rays. The
signal near the GC suffers from large uncertainties in both the dark
matter profile and the astrophysical backgrounds. The isotropic signal
is extremely sensitive to the halo mass function. Some recent analyses
have used these sets of data to constrain dark matter explanations of
cosmic-ray signals, for a variety of dark matter models and spatial
distributions
\cite{Cirelli:2009vg,Cirelli:2009dv,Papucci:2009gd,Chen:2009uq,Hutsi:2010ai}.
In Fig.~\ref{fig:gammas} we show that for the best fit spectra and
spatial distributions in this paper, there is little or no tension
between the predicted total signal and the data in the GC region.

Finally, we also include the following data in order to help regularize
the fit at lower energies and higher energies:
\begin{itemize}
	\item AMS $e^+ +e^-$ cosmic ray spectrum below 10 GeV
	\cite{Alcaraz:2000bf}
	\item HESS $e^+ +e^-$ cosmic ray spectrum above 900 GeV
	\cite{Aharonian:2009ah}
\end{itemize} Fitting to the data from AMS ensures that the background 
models are consistent with the low energy cosmic ray data.

We include systematic errors in our analysis and treat them as
statistical errors because we do not have the full covariance matrix.
The energy calibration error of the Fermi data points is $^{+ 5 \%}_{-
10\%}$, but rather than effectively increasing the error bars, we
allow for freedom in the normalization of the background, discussed in
the next section. The 15$\%$ energy calibration error has been
included in the error bars used for HESS.


\subsection{Solar Modulation \label{sec:solarmod}}

Though our focus is on high energy data, we ensure that our results
are consistent with the low energy (below $\sim$10 GeV) cosmic ray
data from PAMELA and AMS. However, this data is extremely sensitive to
the very local propagation in the heliosphere. To relate the data to
GALPROP predictions for the local interstellar (LIS) spectrum outside
the heliosphere, it is necessary to apply solar modulation corrections
to $e^+ +e^-$ spectra. The solar modulation correction in the
force-field approximation is
\begin{equation}
	\frac{J_\sun(E)}{E^2-m_e^2} =
	\frac{J_{LIS}(E+\Phi)}{(E+\Phi)^2 - m_e^2}
	\label{eq:solarmod}
\end{equation}
where $\Phi$ is the solar modulation parameter and $J$ is the
differential intensity $dn/dE$ \cite{1968ApJ...154.1011G}. Because of
the uncertainty in the force field approximation, we reduce the weight
of the PAMELA and AMS data points below 10 GeV, effectively
multiplying error bars by a factor of 3. This is adequate to stabilize
the fits at low energy.

The solar modulation correction is applied to the GALPROP outputs. We
also use the correction when converting the positron fraction data of
PAMELA into a positron flux data, using the AMS data on the intensity
of $e^+ + e^-$. This will allow the fit to be linear below.

However, these two data sets correspond to different parts of the solar
cycle. We thus apply an \emph{inverse} solar modulation correction
to the AMS data to obtain the unmodulated positron intensity. Denoting
the solar modulation correction by $\hat S_\Phi$, then the positron
signal is obtained from
\begin{align}
	J_{PAM}&(e^+)  = \left( \frac{J(e^+)}{J(e^-) + J(e^+)}
	\right)_{PAM} \times \nn \\ & \hat S_{\Phi^-_{PAM}} \left( \hat
	S^{-1}_{\Phi_{AMS}} \left[ J_{AMS}(e^+ + e^-) \right] \right)
	\label{eq:positronflux}
\end{align}
where $\Phi$'s are solar modulation parameters. $\Phi^-_{PAM}$ is the
solar modulation parameter for the PAMELA electrons, which we allow to
be different from $\Phi^+_{PAM}$ for the positrons. This approximately
captures the charge dependence of the solar physics, visible in the
time-dependent positron fraction at lower energies. In the above
equation we applied $ \hat S_{\Phi^-_{PAM}}$ to the total $e^++e^-$
signal. Because the positrons are at most $\sim 10 \%$ of the total
flux, this approximation is justified. 



\section{Fitting procedure}

We fit for the $\epp$ injection spectrum that, when combined with a
background model, best matches the cosmic ray, gamma ray, and
microwave observations. The steady-state $\epp$ density is linear in
the source function $Q(E,\vec x)$, so we take a Green's function
approach in energy space. The spatial dependence is fixed to be one of 
a few conventional models in each of the cases below.

We inject delta functions of $\epp$ at various energies and compute
the signal from each delta function with GALPROP. Since GALPROP is
discretized, in practice this amounts to propagating an appropriately
normalized bin of energy. For each of these delta functions, GALPROP
computes the steady-state $\epp$ spectrum as well as maps of
synchrotron and IC radiation at various energies.  We solve for the
linear combination of these outputs that best matches the data. The
best-fit injection spectrum solution is simply the same linear
combination of delta function injections (or in our case, energy
bins).

We inject $\epp$ via the source term $Q(E, x)$ in the propagation
equation, Eq.~\ref{eq:diffusion}. For dark matter annihilation, dark
matter decay, and pulsars, the new source function $Q_1(E, \vec x)$ of
both positrons and electrons can be written as
\begin{align}
\label{eq:injection_cases}
Q_1(E,\vec x) = 
\begin{cases}
  \frac{dN}{dE} \ \langle \sigma v \rangle_0 \  BF  \ \frac{ \langle
    \rho_\chi^2 \rangle }{m_\chi^2} \ \frac{f_E}{2} & \text{, ann}\\
  \frac{dN}{dE} \ \tau_\chi^{-1} \ \frac{ \rho_\chi }{m_\chi} \
	    \frac{f_E}{2} & \text{, decay} \\
 \frac{dN}{dE} \ \tau_p^{-1} \ n_p   & \text{, pulsar.}
\end{cases}
\end{align}
  Here $dN/dE$ is the spectrum of electrons or positrons produced per
unit source, normalized such that all the power per unit source goes
into electrons.\footnote{The specific condition can be found in
Sec.~\ref{sec:ann} for dark matter annihilation and
Sec.~\ref{sec:decay} for dark matter decay.}  $\rho_\chi(\vec x)$ and
$m_\chi$ are the energy density and mass of the dark matter. $\langle
\sigma v \rangle_0$ is the thermal freeze-out cross section for
annihilation, $3 \cdot 10^{-26} \text{cm}^3/\text{s}$. $BF$ is a boost
factor (from either particle physics or astrophysics such as
substructure enhancement). $\tau_\chi$ is the lifetime in the case of
dark matter decay. $\tau_p$ and $n_p(\vec x)$ are rate and density
parameters associated with pulsar emission rate and number density.
Finally $f_E = f_E(e^+ + e^-)$ is the fraction of energy going to
electron-positron pairs. If $f_E = 1$, then the total energy of the
electrons will be equal to $m_\chi$ for dark matter annihilation and
$m_\chi/2$ for dark matter decay.

We also consider arbitrary modifications to the energy dependence of
the background primary electron injection,
Eq.~\ref{eq:primary_electrons}. To accomplish this, we include an
extra source of {\it only} electrons which has the same spatial
distribution as the supernovae, Eq.~\ref{eq:source_dist}:
\begin{equation}
  Q_1(E, \vec x) = \frac{dN}{dE} \ \tau_s^{-1} \
         n_s \ \ \ \ \text{, supernova}
	\label{eq:injection_src}
\end{equation} 
where $\tau_s$ is an arbitrary rate parameter that is fixed by
matching to the data.

Because we do not wish to {\it a priori} specify model parameters, we instead
implement the scenarios above with the following electron injection:
\begin{align}
\label{eq:injection}
Q_1(E,\vec x) = 
\begin{cases}
  Q_1(E,\vec x_0) \left( \frac{\rho_\chi(\vec x)}{\rho_\chi(\vec x_0)} \right)^2 & \text{, ann}\\
  Q_1(E,\vec x_0) \left( \frac{\rho_\chi(\vec x)}{\rho_\chi(\vec x_0)} \right) & \text{, decay} \\
  Q_1(E,\vec x_0) \left( \frac{n_p(\vec x)}{n_p(\vec x_0)} \right) & \text{, pulsar} \\
  Q_1(E,\vec x_0) \left( \frac{n_s(\vec x)}{n_s(\vec x_0)} \right) & \text{, supernova.}
\end{cases}
\end{align} 
where the local injection, $Q_1(E,\vec x_0)$ will be determined by the
fit ($\vec x_0$ is our location in the galaxy). The positron injection
is the same, except in the case of the source injection where there
are no positrons injected. Only the spatial profiles distinguish dark
matter annihilation, dark matter decay, or pulsars, in our fits.

We bin the energies of the new source, $Q_1(E,\vec x)$, and treat the
particles in each energy bin independently. For example, we generally
consider the energy range $\sim$ 5-5000 GeV with 17 log spaced
bins. The propagation of a given injection spectrum is just a linear
combination of the propagation of each of the energy bins.

The problem can be treated linearly because high energy $\epp$ are a
tiny perturbation to the matter and radiation of the Galaxy.  High
energy $\epp$ also almost never interact with each other or other
cosmic rays; they dominantly interact with the ISM, radiation, and
magnetic fields.  In GALPROP, the magnetic field is fixed and the
usual feedback between cosmic rays and $B$ field is absent. In this
limit, the propagation of the individual energy bins is independent.

We use this linearity to invert the propagation problem and determine
the injected spectrum $Q_1(E,\vec x_0)$, given some assumptions about
the spatial density of dark matter or pulsars. Define the vector ${\bf
x}$ by $x_i = Q_1(E_i,\vec x_0)$ for energy bin $E_i$. The injection
everywhere else is determined by the assumed spatial
distribution. Also, let $b_j$ be the $j$th data point minus the
galactic background, computed by GALPROP, for that data point.

For each $x_i$, we propagate the injection and obtain a signal
$A_{ji}$.  Thus ${\bf A}$ is a matrix which maps ${\bf x}$ to the
predicted signal, and the columns of ${\bf A}$ give the predicted
signal from each energy bin.  We wish to compare the signal from the
new source, ${\bf A \cdot x}$, with the background-subtracted data,
${\bf b}$.

The best fit ${\bf x}$ is determined by a goodness-of-fit test, which
for a linear problem is a quadratic in the fit parameters:
\begin{equation}
	\chi^2 ={(\bf A \cdot x - b )^T C^{-1} ( A \cdot x - b )}
\end{equation} 
where ${\bf C}$ is a covariance matrix containing the errors on the
data. This is just a quadratic minimization problem. Note that we also
include several other parameters in ${\bf x}$ that allows us to
slightly modify the background predictions and improve the fit. This
is described further in the following subsection.

Finally, it is possible to obtain $dN/dE$ and the other parameters in
Eq.~\ref{eq:injection_cases}. This will be possible for the dark
matter scenarios with additional constraints on $\int E \ dN/dE \ dE$
and where $dN/dE$ cuts off. This is discussed more below when we
describe the results for the scenarios above.


\subsection{Uncertainties \label{sec:uncertainties}}


\begin{figure*}[t]
\begin{center}
\includegraphics[width=\textwidth]{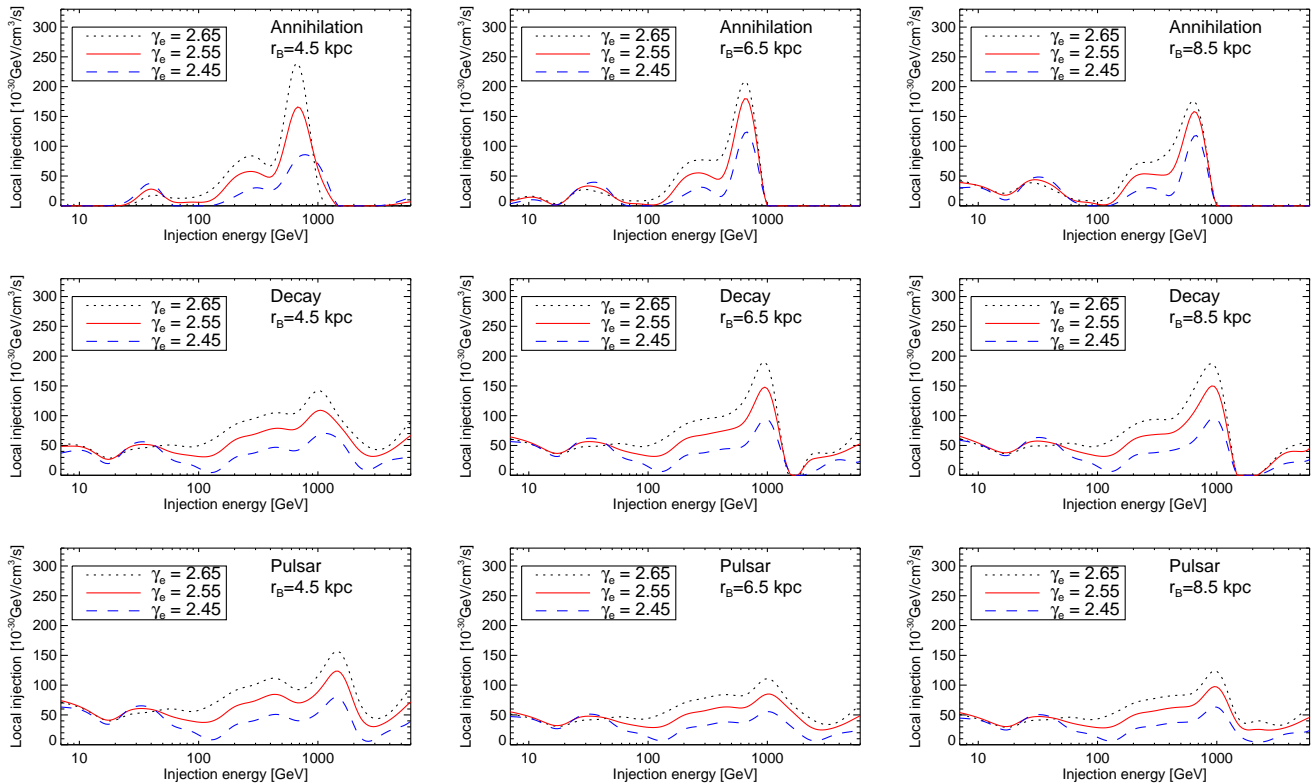}
\end{center}
\caption{All fit results for the three scenarios, over a $3 \times 3$
grid in background electron injection index ($\gamma_e$ = 2.45, 2.55,
2.65) and scale for the magnetic field $r_B = 4.5, 6.5,$ and 8.5
kpc. These spectra were obtained from non-negative fits; the
interpolated local injection density is plotted. Despite a wide range
of assumptions about the background model, the results remain the
same, qualitatively, for each scenario.}
\label{fig:specvar}
\end{figure*}


The predictions for the signals discussed in this paper can have
significant theoretical and astrophysical uncertainty. To capture the
effects of these uncertainties, we include several parameters in the
fits that essentially allow for (small) variations in the background
model. 

The main uncertainties are in the background primary and secondary
$\epp$ injection, since we are fitting all the data that could
possibly constrain this. (For a more detailed discussion of
uncertainties in the cosmic rays signals, see \cite{Delahaye:2010ji,
Simet:2009ne}.)

In the primary injection, we allow $\gamma_e$ in
Eq.~\ref{eq:primary_electrons} to vary in discrete steps. We also fit
for an arbitrary normalization factor $N_p$ relative to the condition
in Eq.~\ref{eq:primary_normalization}. Usually we find $N_p \approx
1.0$ because the condition was chosen to approximately match the Fermi
cosmic ray data.  The source spatial distribution for primary
electrons, Eq.~\ref{eq:source_dist}, is also uncertain. Rather than
considering the full range of possible source distributions, we simply
allow for a different normalization factor of the primary electron
spectrum near the Galactic Center, relative to
Eq.~\ref{eq:primary_normalization}.  Because the diffusion length is
$\sim 1 \kpc$, this will not affect the local cosmic ray signal.  This
normalization factor, $N_h$, is fixed by matching the synchrotron
background prediction onto the Haslam 408 MHz synchrotron map
\cite{1982A&AS...47....1H} for $-90 < b < -5$, averaged over $-10 <
\ell < 10$. The contribution from the new high-energy source is
negligible at this frequency.

\begin{figure*}[phtb]
\begin{center}
(a) \includegraphics[width=.45\textwidth]{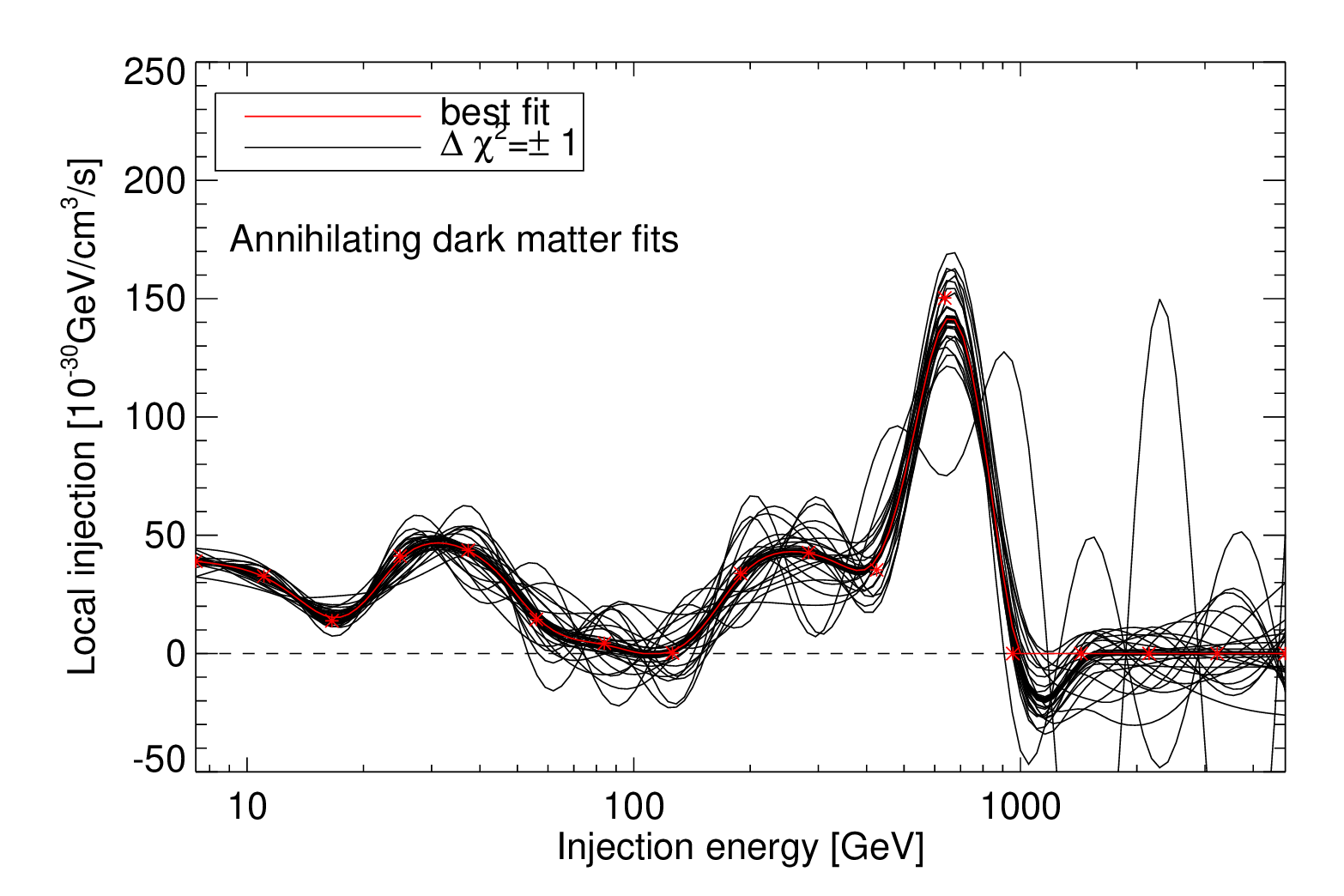}
(b) \includegraphics[width=.45\textwidth]{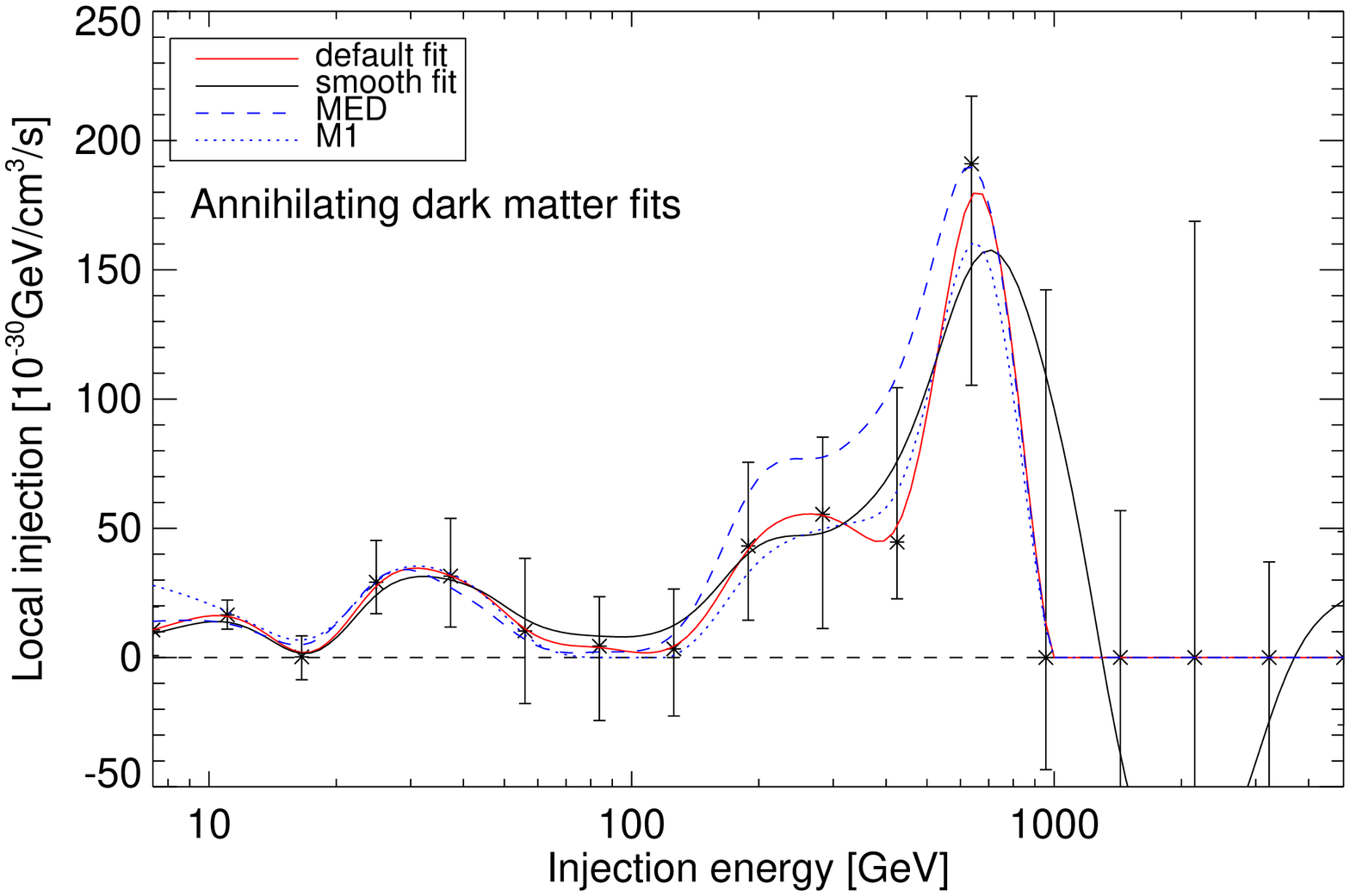} \\
(c) \includegraphics[width=.45\textwidth]{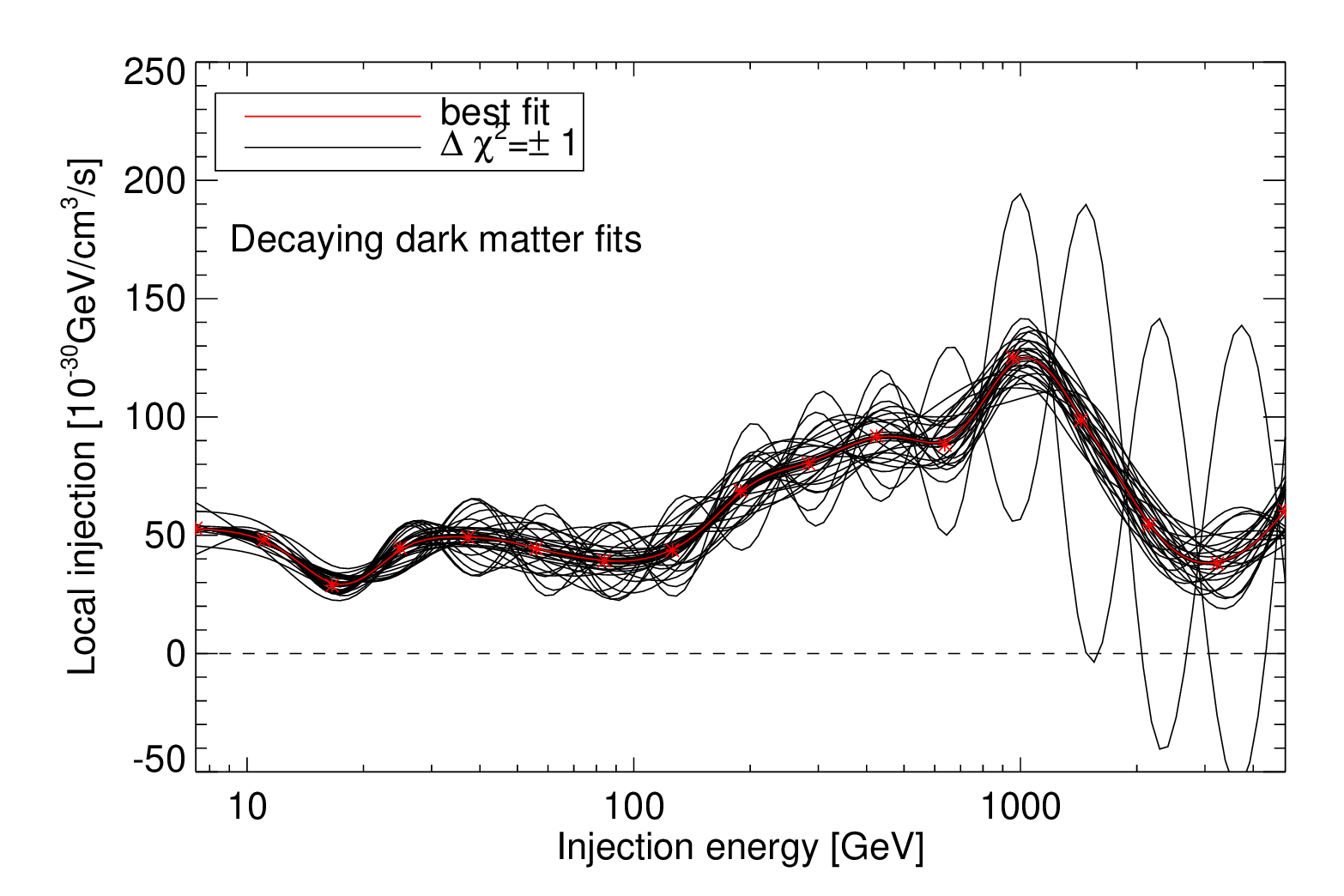}
(d) \includegraphics[width=.45\textwidth]{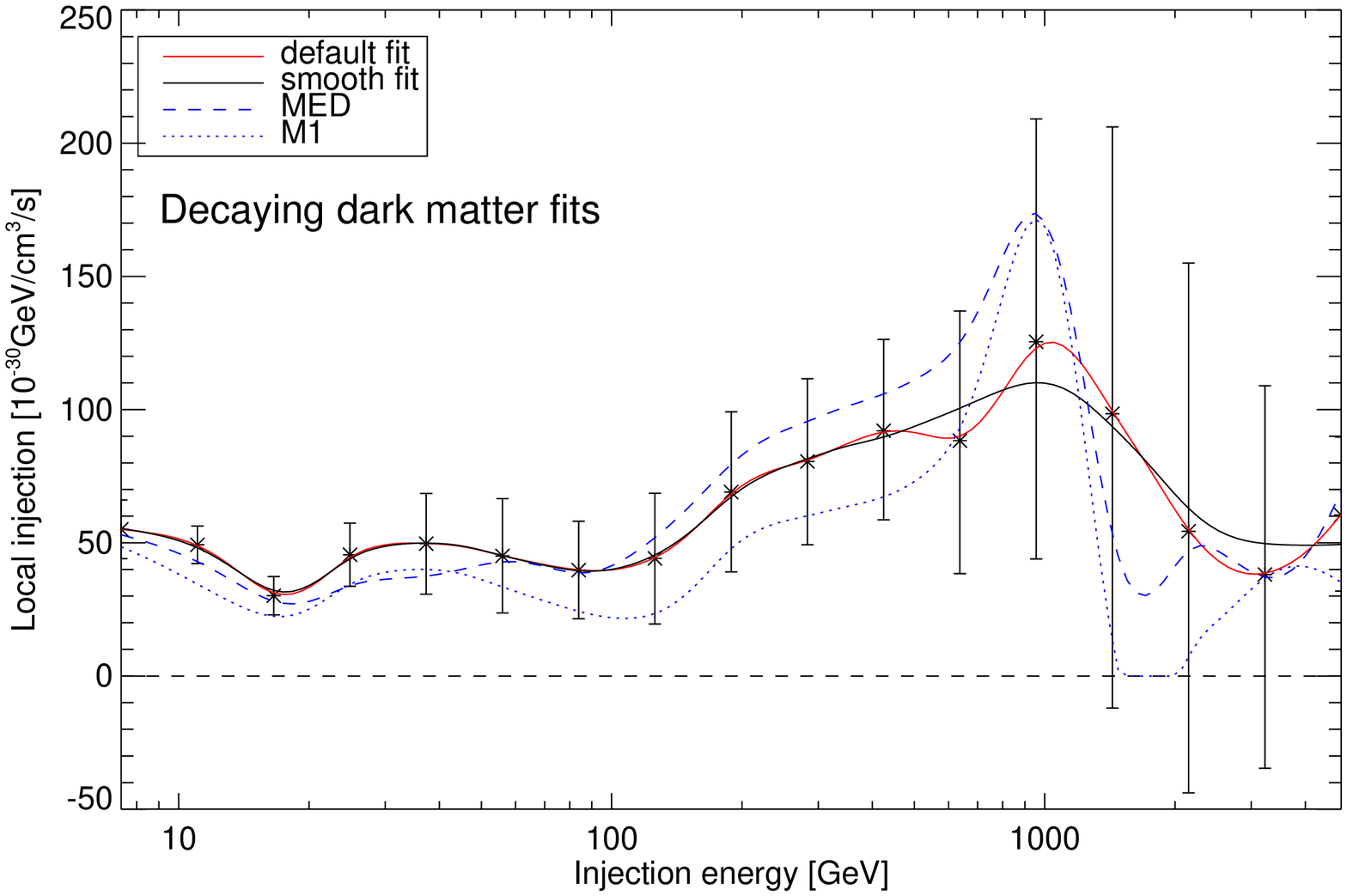}\\
(e) \includegraphics[width=.45\textwidth]{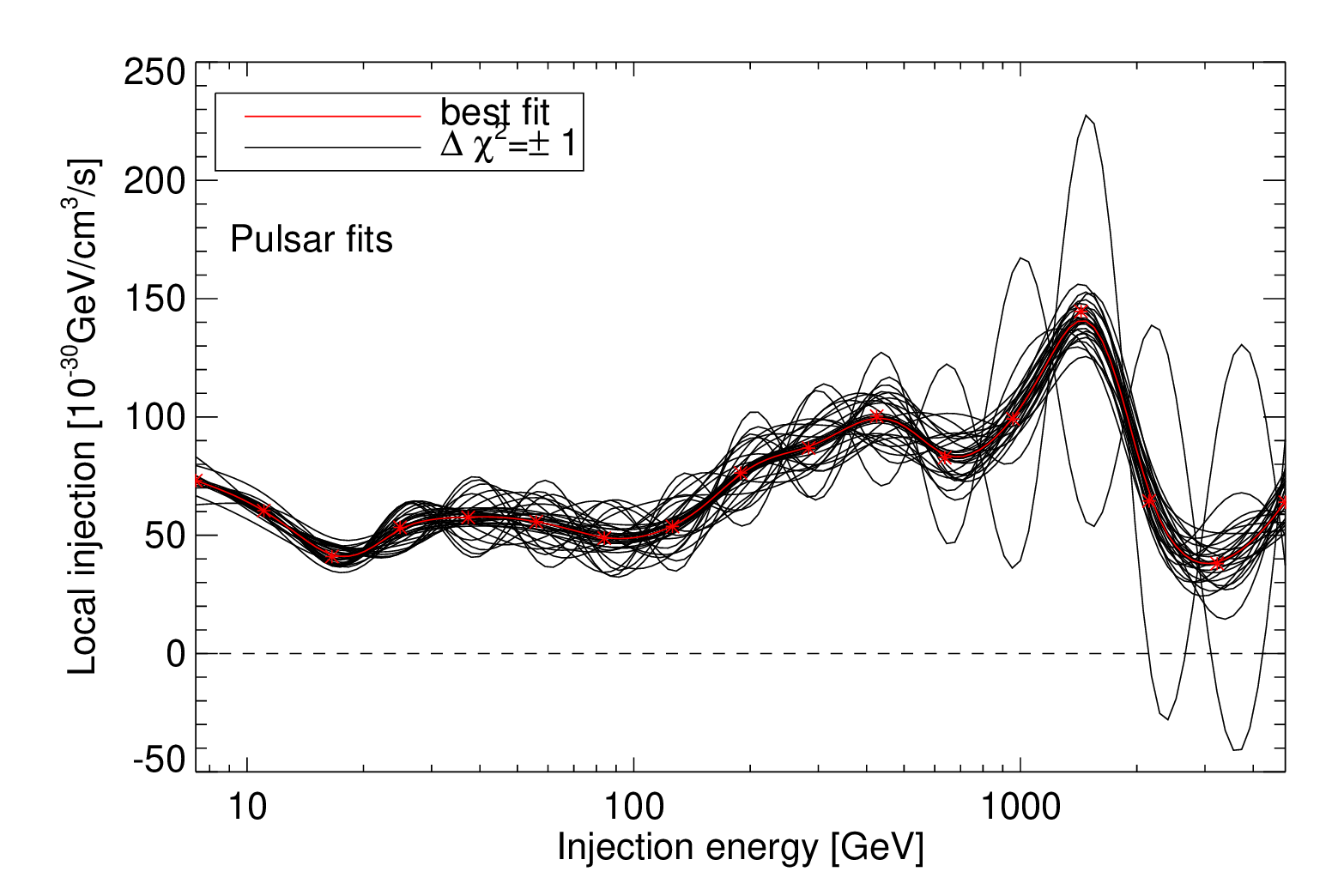}
(f) \includegraphics[width=.45\textwidth]{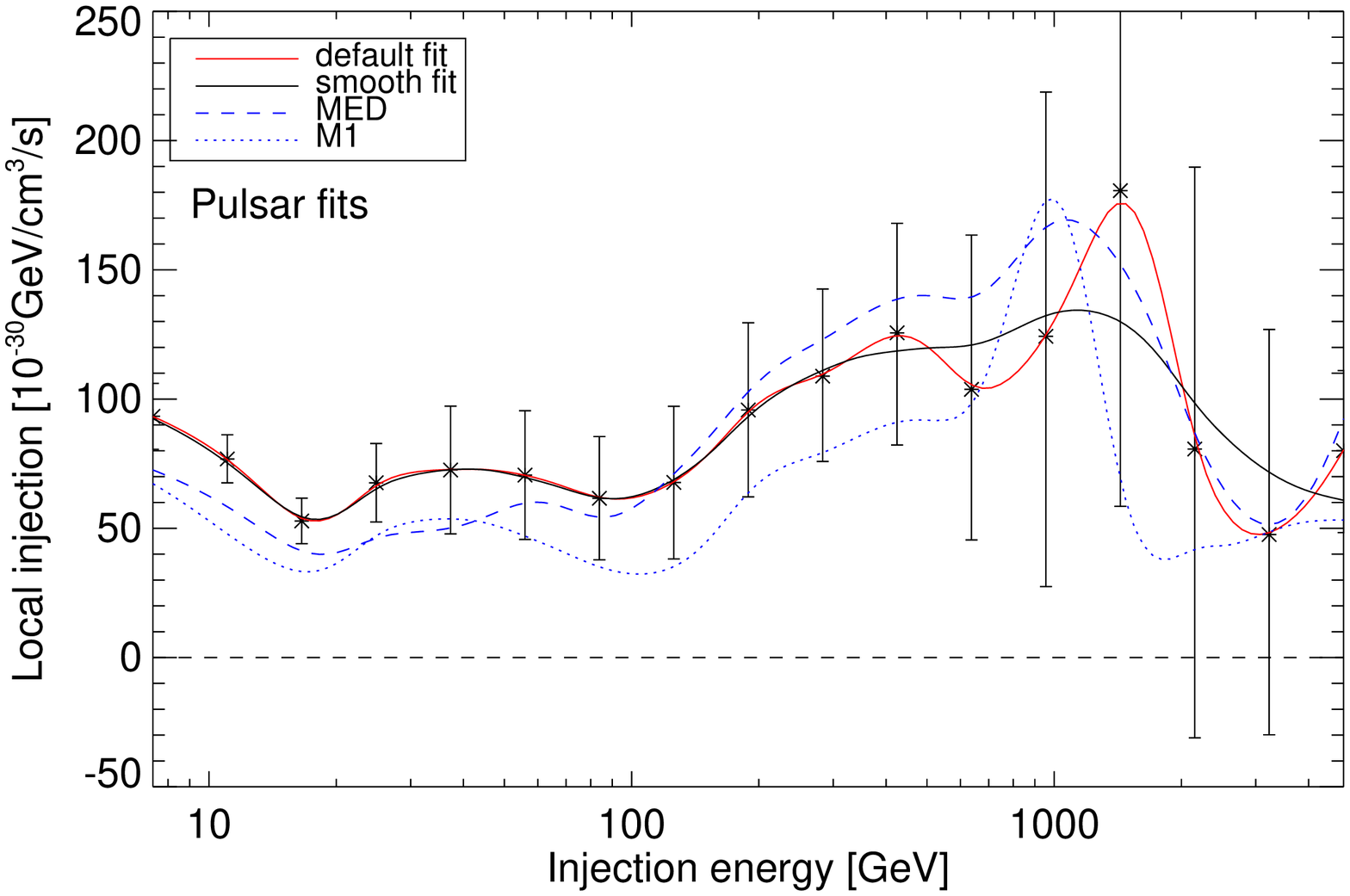}
\end{center}
\caption{In the left column we show the space of possible solutions
within $\Delta \chi^2 = \pm 1$ of the best-fit solution, which was
obtained from a non-negative fit. The red lines are the best fits for
(a) dark matter annihilation, (c) dark matter decay, and (e)
pulsars. The spectra shown are interpolated between the bins (marked
by red stars). In the right column we show other best fits obtained
from using different propagation parameters, given in
Table~\ref{tab:prop}, or a different fitting regulator that enforces
smoothness of the solution (from Eq.~\ref{eq:smoothreg}).  Our results
are robust to very different propagation parameters.  The fit for M2
is not shown because $L$ is only 1 kpc. Since the haze signals extend
out to $\sim$4 kpc or more, it is impossible for this set of
propagation parameters to produce the haze. }
\label{fig:errors}
\end{figure*}

As for the secondary $\epp$, it was shown in \cite{Delahaye:2007fr}
that modifying the propagation parameters effectively changes the
overall normalization of the local steady-state secondary positron
flux by up to an order of magnitude. Thus rather than scanning over a
large set of propagation parameters consistent with all the low energy
cosmic ray data, we allow the normalization of the secondaries to be a
fit parameter, $N_s$.\footnote{We show the effect of changing
propagation parameters on some fits in Fig.~\ref{fig:errors}. The
result does not differ significantly relative to the error bars.}

There are several other adjustable parameters that can improve the fit
and allow for variations in the background model. We give the complete
list below. None of these will change the predictions for other cosmic
ray data.

The following parameters characterize the uncertainties of our
theoretical models. We fit for these parameters simultaneously with
the injection spectrum, as their effects can also be treated linearly:
\begin{itemize}
	\item $N_{IC}$: The normalization of the background IC
	signal near the center of the galaxy, relative to the GALPROP
	prediction. There are many uncertainties in the starlight
	density and spatial variations in the primary electron density
	near the galactic center.
	\item $\Delta S$: Zero-points of the WMAP signal. We allow a
	different one for each data set.
	\item $N_s$: Normalization of secondary local electron
	spectrum, relative to the GALPROP output for our choice of
	propagation parameters. This can vary by up to an order of
	magnitude given theoretical uncertainties
	\cite{Delahaye:2008ua, Simet:2009ne}.
	\item $N_p$: Normalization of {\bf local} primary electron
	spectrum, relative to Eq.~\ref{eq:primary_electrons}. As
	mentioned above, this factor does not have to be the same as
	$N_h$.
\end{itemize} 
We include these parameters in ${\bf x}$, and ${\bf A}$ is enlarged to
include extra columns corresponding to each of these background
signals.

The signals are not linear in the following parameters, so we scan
over a discrete set of these:
\begin{itemize}
	\item $r_B$: The $r$-scale of the galactic $B$ field, where
	the local $B$ field is fixed to 5 $\mu G$. See
	Eq.~\ref{eq:Bfield}. We include $r_B = 4.5, 6.5,$ and $8.5
	\kpc$, corresponding to $B = 33, 18$, and $14 \ \mu G$ in the
	center of the Galaxy. We used $z_B = 2 \kpc$.
	\item $\gamma_e$: The index of the primary electron injection
	spectrum above 4 GeV. We include $\gamma_e = 2.45, 2.50, 2.55,
	2.60, 2.65, $ and $2.70$.
	\item $\Phi_{AMS}, \Phi^+_{PAM},\Phi^-_{PAM}$: Solar
	modulation parameters for AMS and PAMELA, in the force-field
	approximation, as described in Section~\ref{sec:solarmod}.
\end{itemize}
Though we allow these to be fit parameters, clearly in reality they
have some definite form independent of our model. In
Fig.~\ref{fig:specvar} we show the best fit for a grid in $r_B$ and
$\gamma_e$. Though the spectra do change, the qualitative features
remain roughly the same. 


\subsection{$\chi^2$ minimization and regularization \label{sec:reg}}

We are minimizing
\begin{equation}
	\chi^2 ={(\bf A \cdot x - b )^T C^{-1} ( A \cdot x - b )}
\end{equation} 
where ${\bf x}$ is a vector of parameters we fit for, containing the
injection spectrum as well as the normalization parameters and offsets
described above. ${\bf C}$ is a covariance matrix, so it is symmetric
and positive. It can then be shown that the matrix ${\bf A^T C^{-1}
A}$ is positive semi-definite.

Ideally the spectrum we derive is smooth and non-negative. However,
the existence of null (or nearly null) eigenvalues of ${\bf A^T C^{-1}
A}$ means that there are directions in the parameter space where we
can modify the spectrum by large values with little change to the
observed signals. This corresponds to, for example, changing the spectrum
for two neighboring energy bins by a large positive and negative
amount respectively, such that the observed signal remains nearly the
same.

We regularize the spectrum by using only 17 log-spaced bins between 5
GeV and 5000 GeV. We also perform a non-negative quadratic fit following the
algorithm in \cite{Sha02multiplicativeupdates}. All of the fit
parameters should be positive except for $\Delta I$ which we find is
always of the same sign for our data, so we can choose a convention
where it is positive.

To obtain errors on the spectrum, we find the eigenvectors and
eigenvalues of ${\bf A^T C^{-1} A}$. This allows us to change basis
from the parameter space in ${\bf x}$ to a new parameter space ${\bf
y}$ where $\chi^2$ is separately parabolic in each parameter. The
variance of these new parameters ${\bf y}$ is determined by computing
the allowed shift of each parameter, relative to the best fit, such
that $\Delta \chi^2 = \pm 1$. Even though a non-negative constraint
was imposed for the best fit, we consider the entire space of
solutions within $\Delta \chi^2 = \pm 1$.

Each $x_i$ is a linear combination of the $y_i$, so
we sum the squares of the contribution from each $y_i$ to find the
variance in $x_i$. The quoted error on each $x_i$ is the square root
of the variance. Because we are performing a non-negative fit,
however, the positive and negative errors can be different.

In Fig.~\ref{fig:errors} we show the entire range of possible
variations of the best fit injection spectrum with $\Delta \chi^2 =
\pm 1$. We add to the best-fit spectrum all possible variations along
the eigendirections, or all independent variations of $y_i$. (We do
not show the background normalization coefficients and WMAP offsets,
though they are simultaneously varying with the injection spectrum.)

We also considered several alternative methods of regularization,
rather than non-negativity. As an example, we can impose smoothness by
adding terms to $\chi^2$:
\begin{align}
\chi^2_{eff} = \chi^2 & + \ \eta_1 \ {\bf ( D_E \cdot x )^T \cdot (D_E \cdot x)} \nn\\
                   & + \ \eta_2 \ {\bf ( D_E^2 \cdot x )^T \cdot (D_E^2 \cdot x)} 
\label{eq:smoothreg}
\end{align}
where ${\bf D_E}$ and ${\bf D_E^2}$ are finite difference and
second-difference matrices, respectively. These matrices \emph{only}
act on the injection spectrum and not the other fit parameters such as
normalization factors and WMAP offsets. $\eta_1$ and $\eta_2$ are tunable
parameters that control the smoothness of the fit. In
Fig.~\ref{fig:errors} we show the best fit using this regulator
instead of the non-negative regulator above. For an appropriate range
of $\eta_1$ and $\eta_2$ the solution is qualitatively similar to the
non-negative result. Similarly, we tested several other regulation
techniques, such as suppressing variations in nearly null
eigendirections. Again, for ``reasonable" regulators, the result is
qualitatively similar.


\begin{table*}[thb]
\begin{center}
\begin{tabular}{|c|c|c|c|c|c|c|c|} 
\hline 
 & \ SN \ & \ Ann1 \ & \ Ann2 \ & \ Decay \ & \ Pulsars \ & Ann+Pulsar & Ann+SN \\
\hline
Figure	& \ref{fig:fit_source} & \ref{fig:fit_ann} &  & \ref{fig:fit_dec} & \ref{fig:fit_pulsar}  & \ref{fig:fit_ann_puls} & \ref{fig:fit_ann_src}  \\
Einasto $\alpha$ &  & 0.22 & 0.22  & 0.12 & & $0.17^\dagger$ & $0.17^\dagger$ \\
$\gamma_e$	& 2.65  & 2.5 	& 2.5 	& 2.6 	& 2.6 & 2.55 & 2.55 \\
$r_B$ [kpc] 	& 8.5 &  8.5 	& 6.5 	& 4.5 	& 4.5  &  6.5 & 8.5 \\
& & & & & & & \\
$\Phi_{AMS}$ [GeV]      & 0.52 &  0.42 & 0.46   & 0.46  & 0.42  & 0.48 & 0.48 \\
$\Phi_{PAM}^+$  [GeV]	& 0.08 &  0.20 & 0.18	& 0.04 	& 0.02  & 0.12 & 0.18 \\
$\Phi_{PAM}^-$  [GeV] 	& 0.0  & 0.3 & 0.1	& 0.3	& 0.3  & 0.3 & 0.1 \\
& & & & & & & \\
$N_{IC}$ &  1.8 & $1.3$ & 1.6 & $2.5$ & $2.6$ &  $1.3$ & 1.2 \\
$N_s$ & 1.8 & $0.9$ &  1.4 & $0.6$ & $0.5$ & $0.9$ & 1.6 \\
$N_p$ & 1.0 & $1.1$ & 1.1 & $1.0$ & $1.0 $ & $1.0$ &  0.9 \\
& & & & & & & \\
$\chi^2$      & 30* & 139  & 144 & 129 & 148 & 111 & 111 \\
$\chi_{red}^2$ &.51 & .44  & .45 & .41 & .46 & .37 & .37 \\
& & & & & & & \\
$m_\chi$ [GeV] & &  1000 & 1000 & $\gtrsim$16000 &  & 300 & 300 \\
$BF \times f_E$ & & 70 & 70 & & & 10  & 10 \\
$\tau_\chi / f_E$ [s] & & & & $< 10^{26}$ & & & \\
\hline 
\end{tabular}
\end{center}
\caption{Best fit parameters for annihilating dark matter, decaying
dark matter, and pulsar cases to 350 data points. Ann1 and Ann2 had
nearly the same $\chi^2$ but had different $r_B$ so both results are
displayed. In the supernova (SN) injection case there were 91 data points. We
obtained mass, boost factor, and lifetime parameters from the best
fit. In the last two columns we show fit results for linear
combinations of these three scenarios.
The fit errors on the normalization parameters $N$ are less than
5-10$\%$ and thus are not shown. $^\dagger$For the combination cases,
we fixed the dark matter profiles to have Einasto $\alpha = 0.17$.}
\label{tab:results}
\end{table*}


\section{Results}

We determined the best-fit injection spectrum for 350 data points from
Fermi, PAMELA, WMAP, AMS, and HESS. There are 29 fit parameters coming
from 17 energy bins, 3 normalization factors, 6 WMAP offsets, and 3
solar modulation parameters. Including $r_B$ and $\gamma_e$, then
there are 31 fit parameters. Our results are summarized in
Table~\ref{tab:results} and in
Figs.~\ref{fig:fit_ann}-\ref{fig:fit_ann_puls}. The details of the fit
results for each scenario can be found in the following sections.

In each of the following figures, we show the fits to the 
\begin{itemize}
	\item $\epp$ flux data from Fermi, AMS, and HESS
	\item positron flux obtained from combining the AMS and
	        PAMELA data in Eq.~\ref{eq:positronflux}
	\item positron flux fraction $J(e^+)/(J(e^-) +
                J(e^+))$ from PAMELA for comparison, though we did
                not directly fit to this data
	\item WMAP synchrotron data at 23 GHz and 33 GHz, and 23 GHz
	at high $\ell$; the data for 41 GHz and the high $\ell$ data
	for 33 GHz and 41 GHz are included in the fit, but not shown
	because the fit looks extremely similar to the plots already
	shown
	\item Haslam 408 MHz data, used to fix $N_h$, as discussed in
	Sec.~\ref{sec:uncertainties}
	\item Fermi gamma ray data, where the $\pi^0$ background has
	been subtracted \cite{Dobler:2009xz}
\end{itemize} 
along with the best-fit local injection, $E^2 \ Q_1(E,\vec x_0)$.

Before discussing the fits in detail, we emphasize that the results in
Table~\ref{tab:results} and the spectra plotted here are not meant to
be taken as precise answers but as qualitative guidelines for the
necessary spectra, for each scenario, in order to explain the data. As
shown in Fig.~\ref{fig:specvar}, the spectra vary with the background
model, but the general features remain the same. Errors and variations
in the solution were discussed in Section~\ref{sec:reg}. In addition,
the effect of changing propagation parameters is shown in
Fig.~\ref{fig:errors}.

Specific bumps and features in the spectra we find are more likely
signs that the smooth background models we have assumed are not
adequate.  If there is any large systematic or unmodeled effect in the
Fermi cosmic ray data, for example, it can change the features in best
fit spectrum significantly. In particular, note that the shape of the
high energy region of each spectrum above $\sim$100 GeV is only
constrained by the high energy Fermi cosmic ray data since the Fermi
gamma ray data is primarily only sensitive to the total power in this
energy range. The other data are almost completely insensitive to
such high energy particles. Thus the error bars on these bins are
typically the largest. Furthermore, the high energy spectrum is more sensitive
to changes in $\gamma_e$ (see Fig.~\ref{fig:specvar}).

The low energy part of the spectrum is more severely constrained by
all of the data. However, the low energy spectrum is also extremely
sensitive to the bumps and features in the Fermi cosmic ray spectrum
at low energies. This is very likely a sign that some features in the
Fermi cosmic ray spectrum have not been properly included in the
background model. For example, in Fig.~\ref{fig:fit_ann_src} we show a
fit which allows both dark matter annihilation and an arbitrary
modification to the energy dependence of the supernovae-injected
electrons. The low-energy features can be fit by a modification of the
supernova electron spectrum, while dark matter annihilation is still
necessary to explain the signals above 10-20 GeV.


\begin{figure*}[thb]
\begin{center}
\includegraphics[width=.95\textwidth]{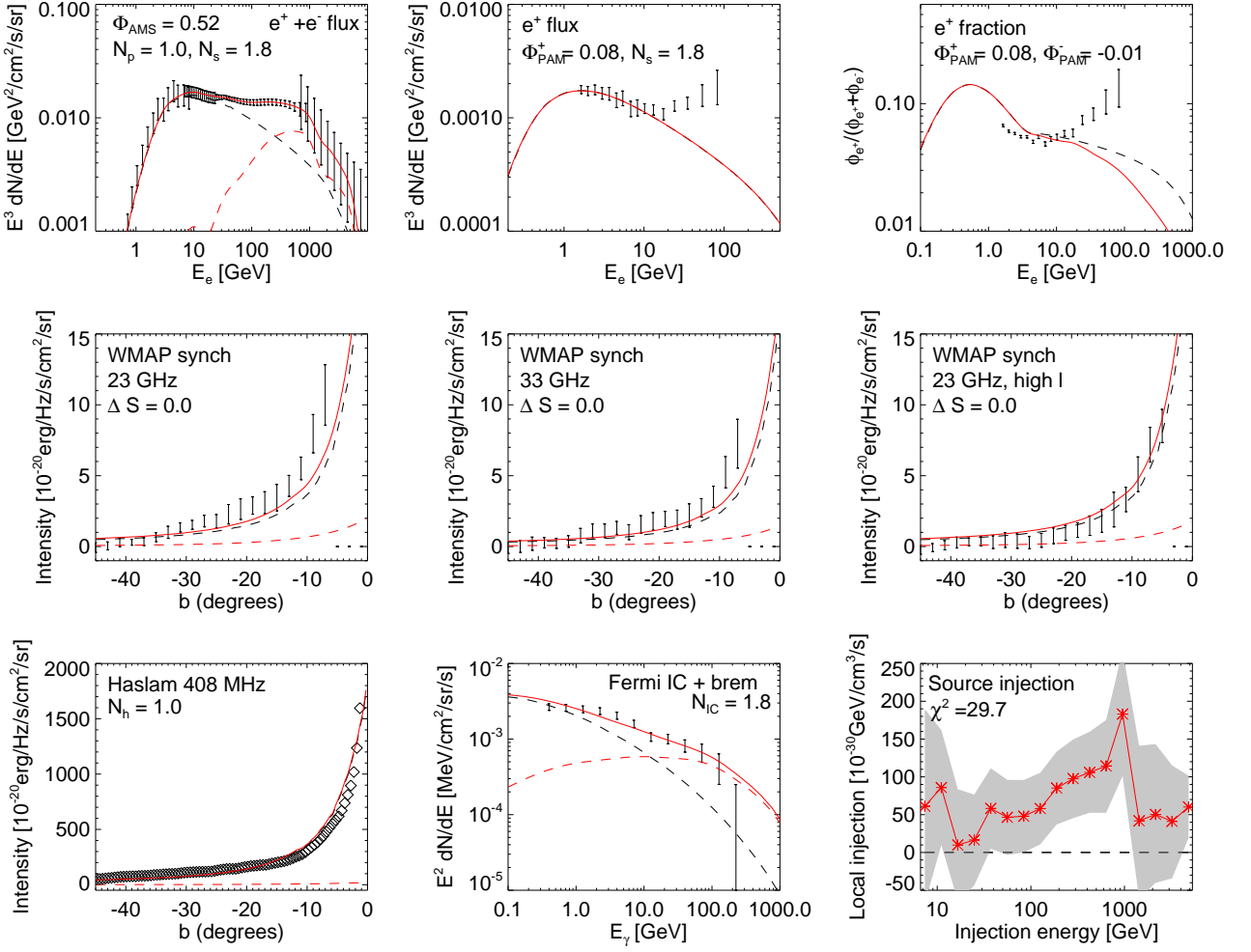}
\end{center}
\caption{Best fit to a modification of the primary electron injection
spectrum, with $\chi^2_{red} \approx .51$.  Black dashed lines are the
background prediction for a model with $\gamma_e = 2.65$ and $r_B =
8.5 \kpc$, though in this case we are fitting for the true
background. Red dashed lines give the contribution of the new source
injection, and solid red lines are the total. The gray shaded region
is the error estimate on the best-fit injection spectrum. We have not
attempted to fit the PAMELA data or the WMAP haze, which are difficult
to produce.}
\label{fig:fit_source}
\end{figure*}

Fig.~\ref{fig:fit_source} shows the ``supernova" fit of the low energy
PAMELA data, all of the $\epp$ data, and the gamma ray data to a disk-like
source with only electrons. This corresponds to a modification of the
background primary electron spectrum and is implemented using the
injection in Eq.~\ref{eq:injection}. The best fit spectrum we found is
a hardening of the injection up to 1 TeV. Though this source
modification can easily match the cosmic ray or IC data, the disk-like
spatial profile and lack of positrons produced are starkly
inconsistent with the synchrotron signal and the PAMELA data. A new
source is required.

\begin{figure*}[t]
\begin{center}
\includegraphics[width=.95\textwidth]{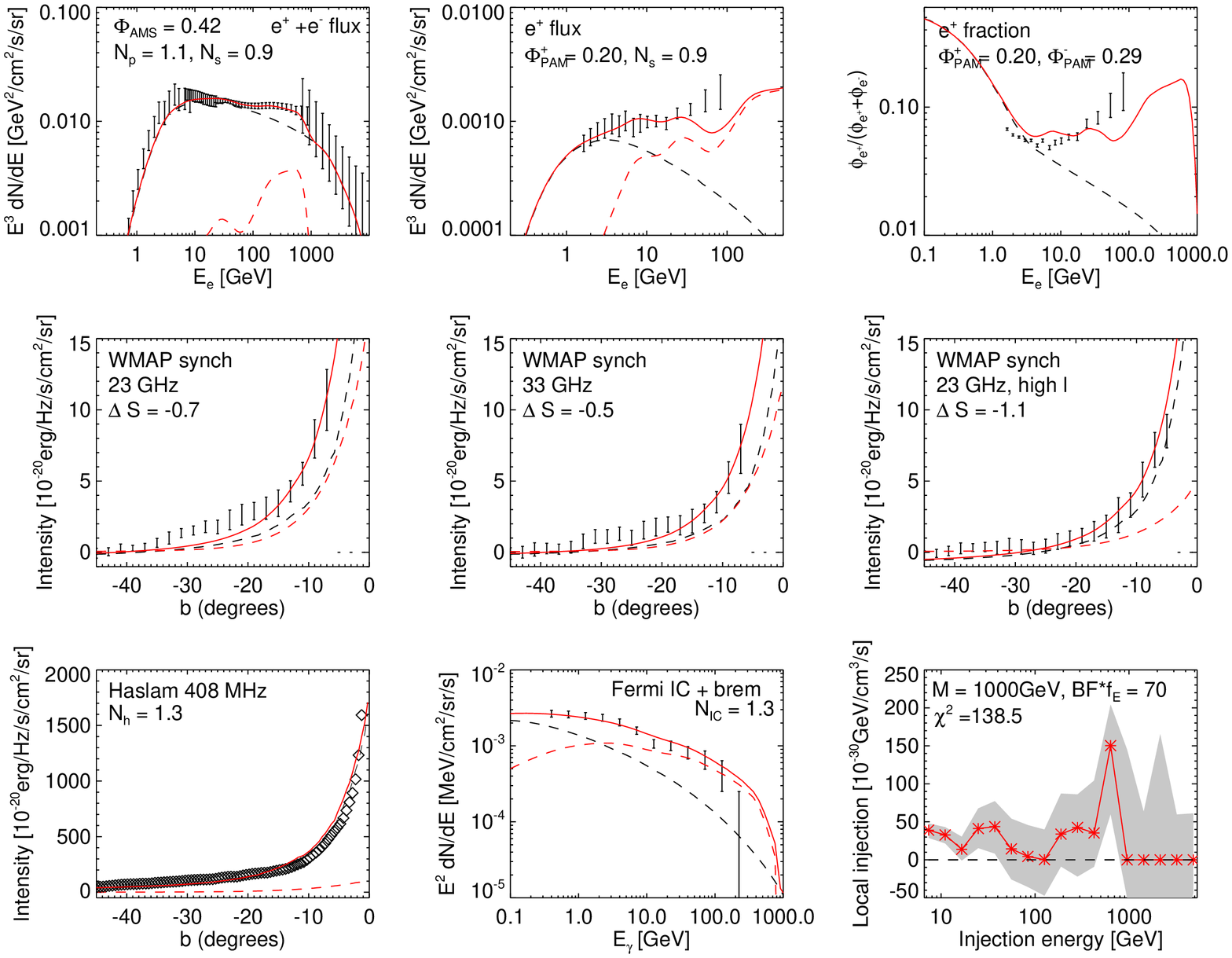}
\end{center}
\caption{Best fit for the annihilating dark matter scenario, with
$\chi^2_{red} \approx .44$. The spatial profile of the dark matter is
Einasto with $\alpha = 0.22$.  Black dashed lines are the background
prediction for a model with $\gamma_e = 2.5$ and $r_B = 8.5 \kpc$. Red
dashed lines give the contribution of the new source injection, and
solid red lines are the total. The gray shaded region is the error
estimate on the best-fit injection spectrum.}
\label{fig:fit_ann}
\end{figure*}


\subsection{Annihilating Dark Matter Results \label{sec:ann}}


The form of the injection for annihilating dark matter was given in
Eq.~\ref{eq:injection} and Eq.~\ref{eq:injection_cases}. We assume the
local dark matter density is $\rho_0 = 0.4$ GeV/$\text{cm}^3$
\cite{Catena:2009mf}.

Conventionally used dark matter halo density profiles are obtained by
simulations and can be approximated by an Einasto profile, with $0.12
\lesssim \alpha \lesssim 0.22$ and $\alpha \approx 0.17$ on average
\cite{Navarro:2003ew}. This does not include substructure effects
which can modify the effective spatial profile used in
Eq.~\ref{eq:injection_cases}, as in \cite{Kuhlen:2009is}.

We allow values of $\alpha = 0.12, 0.17, $ and 0.22, with a core
radius of $r_{-2} = 25 \kpc$. In practice the shallower profile with
$\alpha = 0.22$ is always the best fit to avoid overproducing the
gamma ray signal. These profiles only differ by a factor of $\sim 2$
at $.1 \kpc$ from the center of the galaxy. Though NFW profiles are
also commonly used, their signatures can be approximated by one of
these Einasto profile. We also considered spatial profiles which were
Einasto squared times an $r^{1/4}$ or $r^{1/2}$ scaling, corresponding
to an $r$ dependent cross section \cite{Cholis:2009va}. Using these
profiles can improve the $\chi^2$ by 5-10, but the injection spectrum
does not change significantly.


In the annihilating case we found best fits with magnetic fields of
$r_B =$ 4.5, 6.5, and 8.5 $\kpc$, all with $\chi^2$ around 140
and $\chi_{red}^2 \approx .44$. Conventional magnetic field models
have $r_B$ closer to 8.5 kpc. Furthermore, in this case, the
normalization factors $N$ are $\sim$1, so that the model is
self-consistent. Thus we show the fit with $r_B = 8.5 \kpc$ in
Fig.~\ref{fig:fit_ann}. In Table~\ref{tab:results} we give the fit
parameters for $r_B = 6.5 \kpc$ under the column ``Ann2".

For all three magnetic fields above, we found that an injection index
of $\gamma_e = 2.5$ for the primary electron signal optimized the
ratio between the PAMELA and the Fermi $e^+ + e^-$ apparent dark
matter components. However, for $r_B = 8.5 \kpc$ the fit does not
match the PAMELA data as well, as an excess of cosmic rays above 100
GeV can produce too many gamma rays through IC scattering.

We can estimate several model parameters from the best-fit spectrum by
relating Eq.~\ref{eq:injection} and Eq.~\ref{eq:injection_cases}. To
find the dark matter mass, we assume $dN/dE$ cuts off at around
$m_\chi$. Though this estimate of $m_\chi$ depends on the rather
uncertain high-energy part of the injection spectrum, values of
roughly 1 TeV are expected given the turnover in the $e^+ +
e^-$ data around 600-1000 GeV and the turnover in the gamma-ray spectrum at
100-200 GeV.

Next, $dN/dE$ was defined such that that the total energy of the
emitted particles sums to the mass of dark matter:
\begin{equation}
  \int E \frac{dN}{dE} dE = m_\chi.
\end{equation} 
Therefore, integrating the local injection multiplied by energy gives
\begin{align}
	\int E \  Q_1(E,\vec x_0) dE = 
	     \langle \sigma v \rangle_0 \ BF \
	    \frac{ (\rho_0)^2 }{m_\chi} \ \frac{f_E(e^++e^-)}{2}.
\end{align} 
Given an estimate of $m_\chi$, we can therefore estimate $BF \times
f_E(e^+ + e^-)$ in terms of the best fit local injection and known
parameters. 


\begin{figure*}[t]
\begin{center}
\includegraphics[width=.95\textwidth]{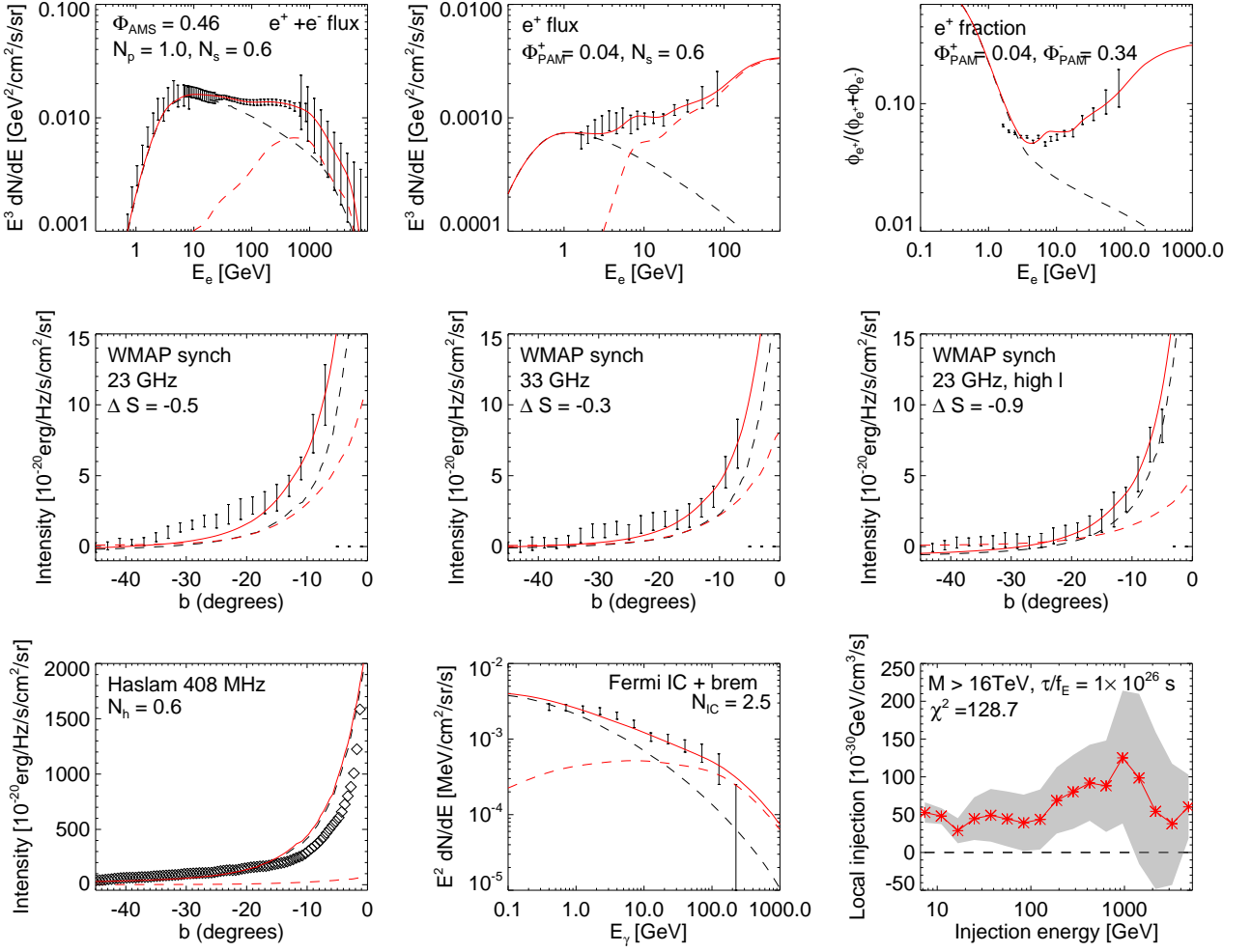}
\end{center}
\caption{Best fit for the decaying dark matter scenario, with
$\chi^2_{red} \approx .41$. The spatial profile of the dark matter is
Einasto with $\alpha = 0.12$. Black dashed lines are the background
prediction for a model with $\gamma_e = 2.6$ and $r_B = 4.5 \kpc$. Red
dashed lines give the contribution of the new source injection, and
solid red lines are the total. The gray shaded region is the error
estimate on the best-fit injection spectrum.}
\label{fig:fit_dec}
\end{figure*}



\begin{figure*}[t]
\begin{center}
\includegraphics[width=.95\textwidth]{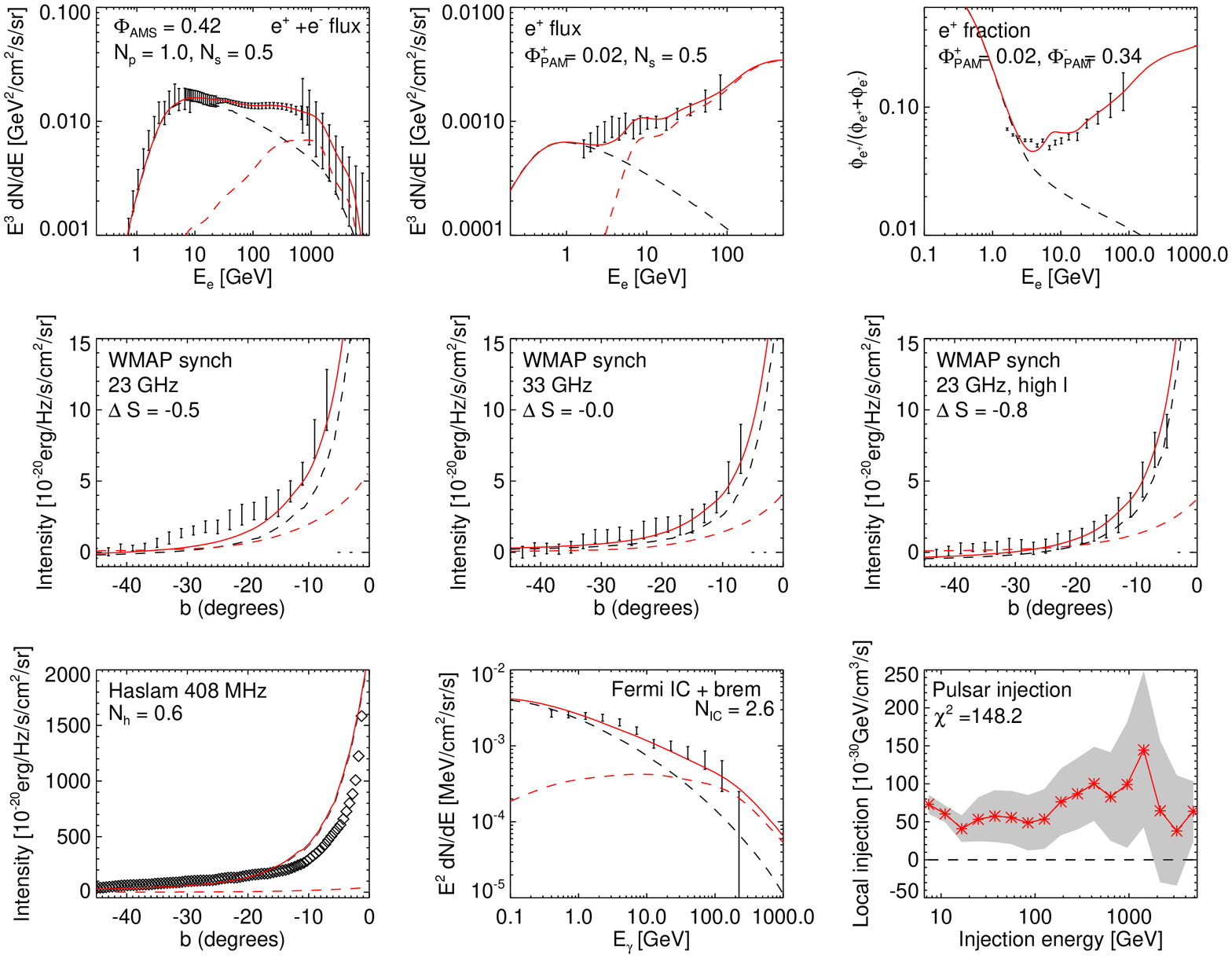}
\end{center}
\caption{Best fit for the pulsar scenario, with $\chi^2_{red} \approx
.46$.  The pulsar profile is given by
Eq.~\ref{eq:pulsarprofile}. Black dashed lines are the background
prediction for a model with $\gamma_e = 2.6$ and $r_B = 4.5 \kpc$. Red
dashed lines give the contribution of the new source injection, and
solid red lines are the total. The gray shaded region is the error
estimate on the best-fit injection spectrum. }
\label{fig:fit_pulsar}
\end{figure*}


\subsection{Decaying Dark Matter Results \label{sec:decay}}

For the decaying dark matter case, we assume the same range of dark
matter density profiles as in the annihilating case. Again, in
practice we will be limited to the case where $\alpha = 0.12$. This
time a steeper profile is required to produce sufficient synchrotron
signal to fit the WMAP data.

The model parameters can be determined from
Eq.~\ref{eq:injection_cases} and Eq.~\ref{eq:injection}. We assume
$dN/dE$ cuts off at around $m_\chi/2$ this time. Again, this cutoff is
rather sensitive to the high-energy part of the spectrum, which has
large error bars, but values of $\gtrsim 2\TeV$ are expected given the
data.

By definition, $dN/dE$ satisfies
\begin{equation}
  \int E \frac{dN}{dE} \ dE = m_\chi/2.
\end{equation} 
Again, we integrate the local injection multiplied by energy, giving
\begin{equation}
	\int E  \ Q_1(E, \vec x_0)\  dE =  
    \  \tau_\chi^{-1} \ \frac{ \rho_0}{2} \ \frac{f_E(e^+ + e^-)}{2}.
\end{equation} 
This allows us to determine the dark matter lifetime over the
energy fraction. However, note that in many cases, $dN/dE$ does not cut
off in the energy ranges we consider and the spectrum is essentially
unconstrained at higher energies. Then we only obtain bounds on the
mass and lifetime.

The best fit is shown in Fig.~\ref{fig:fit_dec}. There is no clear
mass cutoff in the best-fit spectrum, so the mass of the particle can
be from $\sim 4 \TeV$ to greater than 16 TeV. 

Because in the decaying scenario the injected power is proportional to
$\rho_\chi$ and not $\rho_\chi^2$, generally it is harder to generate
enough synchrotron and IC signal. Both of these signals are in regions
at least 5 degrees off of the galactic plane. The steeper dark matter
profile with $\alpha = 0.12$ is not enough to produce the signals.

We found $r_B = 4.5 \kpc$ can increase synchrotron near the center of
the galaxy, but this gives a somewhat unconventionally high value of
the magnetic field in the GC, $33 \mu G$. Fig.~\ref{fig:haslam} shows
that $r_B = 4.5 \kpc$ also gives the poorest fit to the Haslam data,
especially compared to $r_B = 8.5 \kpc$. In addition, a somewhat large
injection of low energy electrons and positrons is required. However,
for this large magnetic field, the IC signal drops. Thus the
normalization $N_{IC}$ is rather large, $N_{IC} \sim 2.4$. Even for
fits with $r_B = 6.5 \kpc$, it was necessary for $N_{IC} \sim 2$ to
obtain sufficient IC signal. This corresponds to rather high starlight
density.

While it is possible that the decaying dark matter can also produce
gamma rays directly or through FSR, these signals are typically at
higher energies, above 10-100 GeV. In this case, the large $N_{IC}$
factor for the background IC signal indicates that there is a
depletion of gamma rays at low energies, below 10 GeV.

Though the decaying scenario nominally gives the best $\chi^2 \approx
130$ and $\chi_{red}^2 \approx .41$, the large normalization factors
demand a more self-consistent modeling of backgrounds and
uncertainties in order to fully justify the goodness of fit.


\begin{figure*}[t]
\begin{center}
\includegraphics[width=.95\textwidth]{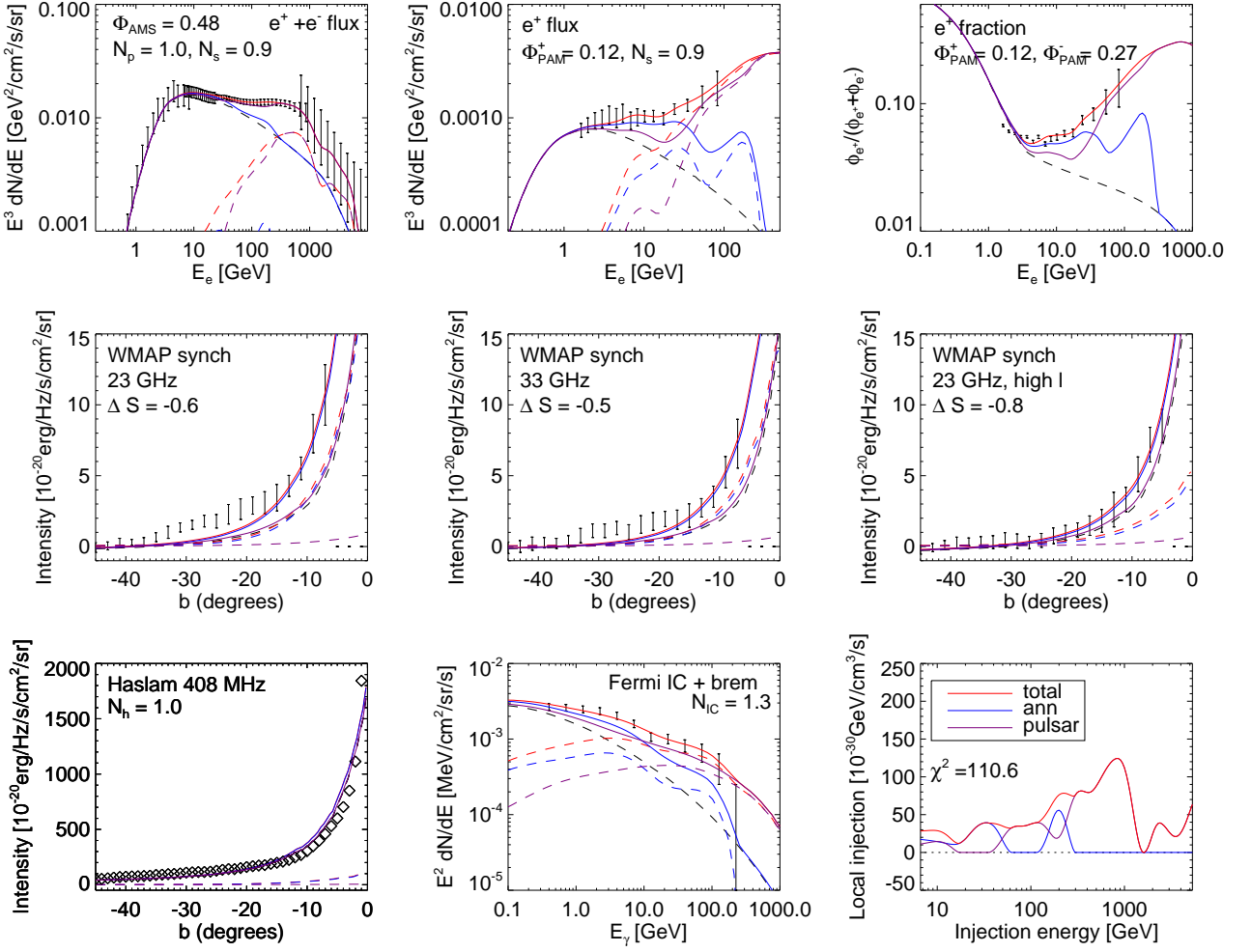}
\end{center}
\caption{Best fit for a linear combination of annihilating dark matter
and pulsars, with $\chi^2_{red} \approx .37$. The dark matter spatial
profile is Einasto with $\alpha = 0.17$. Black dashed lines are the
background prediction for a model with $\gamma_e = 2.65$ and $r_B =
8.5 \kpc$. Colored dashed lines give the contributions from new
sources, while colored solid lines give the total signal from new
source and background.}
\label{fig:fit_ann_puls}
\end{figure*}


\subsection{Pulsar Results}

The range of types of pulsars, their spatial distributions, and their
$\epp$ spectra is not well determined. As a crude model, we posit some
spatial profile for the number density of pulsars and assume the
spectrum of electrons and positrons has the same energy dependence
everywhere. Generally pulsars are concentrated in the galactic disk, 
making it difficult to produce the haze. In Fig.~\ref{fig:haze} we
compare the morphology of the synchrotron haze produced by pulsars to
that produced by dark matter. It is possible, however, that certain
types of pulsars have a more spherical distribution
\cite{Malyshev:2010xc}.

We consider the following range of pulsar profiles, which span those
typically used in the literature. (See \cite{Harding:2009ye,
FaucherGiguere:2009df} for examples and references.) Assume a density
profile of the form
\begin{equation}
	n_p \propto \exp \left( - \frac{ |z| }{z_p} \right) n_\rho (\rho)
	\label{eq:pulsarprofile}
\end{equation}
where $\rho = \sqrt{x^2 + y^2}$ and the origin is at the center of the
galaxy. We include profiles with $z_p = 0.08, 0.5,$ and 2 kpc. For
the radial profile,
\begin{eqnarray}
   n_{\rho} \propto \exp \left( - \frac{\rho}{4.5 \kpc} \right).
\end{eqnarray} 
In practice, the three cases above look nearly identical because of
diffusion. Another commonly used profile has $n_{\rho} \sim \rho \exp
\left( - \rho/4.0 \kpc \right)$. However we do not consider this
option further because the suppressed density near the center of the
galaxy makes it even more difficult to produce the haze.

Qualitatively, the pulsar results, Fig.~\ref{fig:fit_pulsar}, are
rather similar to the decaying dark matter results, though the fits
are even worse because of the disk-like rather than spherical
profile. The best fit has $r_B = 4.5 \kpc$ with significant low-energy
injection and large normalization factor $N_{IC}$ of 2.6. Though it is
possible that pulsars can produce many low-energy gamma rays, it is
unlikely these gamma rays can compensate for the background gamma-ray
signal being $2-3$ times too low.  For $r_B = 6.5 \kpc$ or $8.5 \kpc$,
the pulsar scenario cannot produce sufficient synchrotron signal.


\begin{figure*}[t]
\begin{center}
\includegraphics[width=.95\textwidth]{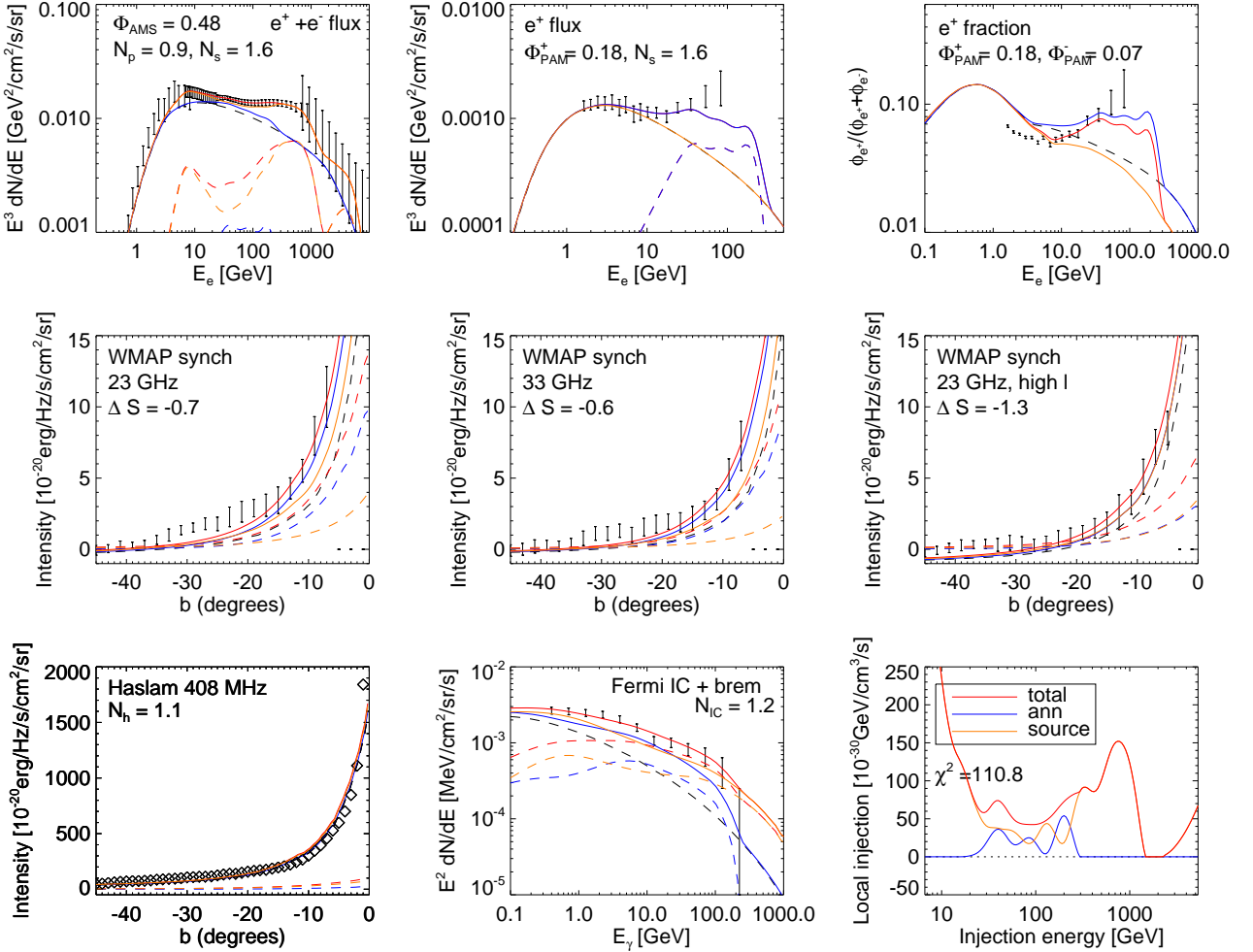}
\end{center}
\caption{Best fit for a linear combination of annihilating dark matter
and an arbitrary modification to the primary electron injection
spectrum, with $\chi^2_{red} \approx .37$. The dark matter spatial
profile is Einasto with $\alpha = 0.17$. Black dashed lines are the
background prediction for a model with $\gamma_e = 2.55$ and $r_B =
8.5 \kpc$. Colored dashed lines give the contributions from new
sources, while colored solid lines give the total signal from new
source and background.}
\label{fig:fit_ann_src}
\end{figure*}


\subsection{Combination Results}

We fit for linear combinations of annihilation, decay, pulsars, and
supernova injections, which not surprisingly can provide better fits and
alleviate the problems of each individual scenario. However, this
extra freedom means that fits are much less constrained. For the fits
presented here, errors in the spectra are much larger the spectra
themselves and are not shown.

To avoid the large normalization factors necessary for dark matter
decay or pulsars to fit the data, we only consider linear combinations
that include dark matter annihilation. Thus enough synchrotron can
easily be produced with lower magnetic fields. We also fix $\alpha$ to
the middle value of $\alpha = 0.17$.

Generally, the power from dark matter annihilation at 10-100 GeV
produces sufficient synchrotron for a large fraction of the WMAP
signal, while the power at 100-5000 GeV produces part of the gamma ray
signal. Decaying dark matter or pulsars can then freely produce the
rest of the local cosmic ray signal without significantly impacting
the predictions from near the galactic center. Because decaying dark
matter and pulsars do not give very different answers we don't show
the fit to all three.

Fig.~\ref{fig:fit_ann_puls} is the fit to annihilation and
pulsars. The fit to annihilation and decay is very similar, and thus
not shown. Fig.~\ref{fig:fit_ann_src} is the fit to annihilation and a
modification of the background primary electrons. This case is perhaps
the most realistic of the possibilities, if a new source is
allowed. There is an unusual spike at low energies which just allows
the primary electrons to more precisely match the features in the Fermi
low energy cosmic ray data, without overproducing any
other signal. We have not shown a fit to a combination of pulsars and
a modification of the primary electrons because, as mentioned above,
for both scenarios it was extremely difficult to produce the haze. A
linear combination of the two would not alleviate this problem.


\section{Conclusions}

We have thoroughly examined the annihilating dark matter, decaying
dark matter, and pulsar explanations of the recent anomalous cosmic
ray, gamma ray, and synchrotron signals. We investigated whether each
scenario can fit all of these data simultaneously. Our analysis is
independent of the particle physics or pulsar model details of each
scenario and only depends on the spatial profiles and background
models. We determined the necessary injection spectrum of electrons
and positrons in each case in order to reproduce the data, including
the effects of different background models, propagation models, and
solar modulation. 

Though decaying dark matter is the best fit, the large normalization
factors suggest that it will be difficult to find a fully
self-consistent model with decaying dark matter that can explain the
data, without changing some aspect of our model by a large amount. In
particular, it may be necessary to find either a radiation field
model with roughly twice as much starlight to produce enough IC, a
rather large magnetic field of $33 \mu G$ in the Galactic center,
enormous amounts of low energy electrons or gamma rays injected, a
much steeper dark matter profile, or a combination of these.

Pulsars give the worst fit; the disk-like profile makes it nearly
impossible to produce both the gamma ray and synchrotron signals. Much
like the decaying dark matter case, this suggests that dramatic
re-assessments of backgrounds and models are necessary to find a
self-consistent interpretation of the data.

Annihilating dark matter, however, has self-consistent fits with
conventional astrophysical background models. Though we had to choose
a somewhat shallower dark matter halo profile with Einasto $\alpha =
0.22$, it is still within the current range of profiles found by
simulations. Furthermore, we can satisfy the gamma ray constraints
from the GC. The boost factors are $\sim 70/f_E$, which at first seems
much lower than the boost factors of $\sim 1000$ often used in the
literature. Several factors enter in this difference: our use of the
updated $\rho_0 = .4$ GeV/cm$^3$ rather than $\rho_0 = .3$ GeV/cm$^3$
\cite{Catena:2009mf}, the relatively hard spectrum allowed by the fit,
and our assumption that the cutoff of the spectrum is $m_\chi$.  Given
these factors, our result of $\sim 70/f_E$ is typical of the models
discussed in the introduction of this paper.  However, the shape of
the spectrum, combined with a lack of $\pi^0$ or $\bar p$ production,
may still be difficult to achieve in current particle physics models
of dark matter.

Our results should be regarded as qualitative guidelines to injection
spectra. The specifics will necessarily change as both Galactic models
and data are refined. The WMAP ``haze'' data will be superseded by
data from Planck \cite{Planck:2006uk}, while data from Fermi and
PAMELA will improve. In addition, cosmic ray data from AMS-02
\cite{AMS02} may also soon be available.  If the data does not change
substantially, and if current models indeed describe Galactic
propagation and interactions, then the qualitative results of this
paper will remain valid.

\begin{acknowledgments}
We are grateful to Ilias Cholis, Marco Cirelli, Tracy Slatyer, and
Neal Weiner for helpful discussions and comments on this paper. The
computations in this paper were run on the Odyssey cluster supported
by the FAS Research Computing Group at Harvard.  TL is supported by an
NSF graduate fellowship.  DPF is partially supported by NASA Theory
grant NNX10AD85G.  This research made use of the NASA Astrophysics
Data System (ADS) and the IDL Astronomy User's Library at
Goddard.\footnote{Available at \texttt{http://idlastro.gsfc.nasa.gov}}
\end{acknowledgments}

\bibliographystyle{apsrev4-1}

\bibliography{spectrum}

\end{document}